\documentclass{article}

\usepackage{arxiv}
\usepackage{url}
\usepackage{cite}
\usepackage{amsmath,amssymb,amsfonts}
\usepackage{algorithmic}
\usepackage{graphicx}
\usepackage{textcomp}
\usepackage{xcolor}
\usepackage{enumitem}
\usepackage{diagbox}
\usepackage{multirow}
\usepackage{listings}
\usepackage{bigstrut}
\usepackage{threeparttable}
\usepackage[linesnumbered,ruled,vlined]{algorithm2e}
\usepackage{booktabs}
\usepackage{bm}
\usepackage{arydshln}
\usepackage{framed}
\usepackage{amssymb}
\usepackage{subcaption}
\usepackage{hyperref}
\usepackage{pifont}
\usepackage{tikz}
\usepackage{graphicx}
\usetikzlibrary{shapes.geometric}
\usepackage{cite}
\usepackage[numbers,sort&compress]{natbib}
\usepackage[affil-it]{authblk}

\newcommand*{\Bigcircle}{\tikz[]{\raisebox{-.6mm}{\node[shape=circle,draw]{};}}}
\newcommand*{\Bigbullet}{\tikz[]{\raisebox{-.6mm}{\node[shape=circle,draw, fill=black, text=white]{};}}}
\newcommand*{\Bigtriangle}{\raisebox{-.6mm}{\scalebox{1.3}{$\triangle$}}}

\newcommand*{\circled}[1]{\tikz[baseline=(char.base)]{
		\raisebox{0mm}{\node[shape=circle,draw,inner sep=0.5pt] (char) {\footnotesize #1};}}}

\newcommand*{\encircled}[1]{\tikz[baseline=(char.base)]{
		\raisebox{0mm}{\node[shape=circle,draw,inner sep=0.5pt, fill=black, text=white] (char) {\footnotesize #1};}}}

\newcommand{\trinum}[1]{\large {\mathpalette\dotrinum{#1}}}
\newcommand{\dotrinum}[2]{{%
		\vphantom{\triangle}%
		\ooalign{%
			$#1\triangle$\cr\hidewidth\scaleraise{$#1#2$}\hidewidth\cr
		}%
}}
\newcommand{\scaleraise}[1]{%
	\raisebox{.3\height}{\scalebox{0.5}{#1}}%
}

\newcommand{\tab}{\hspace*{1em}}
 
\newcommand{\code}[1]{{\fontfamily{cmtt}\fontseries{m}\fontshape{n}\selectfont\small{#1}}}
\newcommand{\pricemani}{price manipulation}
\newcommand{\tool}{\textsc{DeFiRanger}}
\newcommand{\defi}{DeFi}

\def\arxiv{1}

\begin{document}
	
	\title{\tool{}: Detecting Price Manipulation Attacks on DeFi Applications}
	\newcommand\CoAuthorMark{\footnotemark[\arabic{footnote}]}
	
	\author[1]{Siwei Wu}
	\author[1]{Dabao Wang}
	\author[1]{Jianting He}
	\author[1]{Yajin Zhou\thanks{Corresponding author (yajin\_zhou@zju.edu.cn).}\hspace*{0.4em}}
	\author[1]{Lei Wu}
	\author[2]{Xingliang Yuan}
	\author[1]{Qinming He}
	\author[1]{Kui Ren}
	\affil[1]{Zhejiang University}
	\affil[2]{Monash University}

	\maketitle
	
	\begin{abstract}

The rapid growth of Decentralized Finance (DeFi) boosts
the Ethereum ecosystem.
At the same time, attacks towards DeFi applications (apps) are increasing.
However, to the best of our knowledge, existing smart contract
vulnerability detection tools cannot be directly used
to detect DeFi attacks. That's because they lack the capability to
recover and understand high-level
DeFi semantics, e.g., a user trades a token pair \code{X} and \code{Y}
in a Decentralized EXchange (DEX).

In this work, we focus on the detection of
two types of new attacks on DeFi apps, including
\textit{direct} and \textit{indirect} price manipulation attacks.
The former one means that an attacker \textit{directly}
manipulates the token price
in DEX by performing an unwanted
trade in the same DEX by attacking the vulnerable DeFi app.
The latter one means that an attacker \textit{indirectly} manipulates
the token price of the vulnerable DeFi app (e.g., a lending app).
To this end, we propose a platform-independent way to
recover high-level DeFi semantics by first constructing
the cash flow tree from raw Ethereum transactions and then lifting the
low-level semantics to high-level ones, including token trade,
liquidity mining, and liquidity cancel. Finally, we detect price manipulation attacks
using the patterns expressed with the recovered DeFi semantics.

We have implemented a prototype named \tool{} and applied it to
more than $350$ million transactions. It successfully
detected  $432$ real-world attacks in the wild.
We confirm that
they belong to four known security incidents and five zero-day
ones. We reported our findings. Two CVEs have been assigned.
We further performed an attack analysis to reveal the root cause of
the vulnerability, the attack footprint, and the impact of the attack.
Our work urges the need to secure the DeFi ecosystem.

\end{abstract}

	\section{Introduction}
\label{sec:introduction}

The recent Decentralized Finance (\defi{}) boom brings Ethereum a new climax,
attracting $28$ billion USD locked in \defi{} apps up to $31$st January 2021.
With the rapid growth of the \defi{} ecosystem, security issues are also emerging.
\defi{-related} security issues have been reported, including
front-running~\cite{eskandari2019sok,daian2020flash,zhou2020highfrequency}, 
Pump-and-Dump (P\&D) scams~\cite{kamps2018moon,xu2019anatomy},
and flash loan attacks~\cite{qin2020attacking}.
In addition, code and logic vulnerabilities in DeFi apps bring many security 
incidents~\cite{pickle,88mph,origin,akropolis,Opyn,UniswapAndLendf.Me,bZx-hack-I,bZx-hack-II,WarpFinance-incident,
Cheese-bank-incident,Value-incident,Harvest-hack}.

Existing detection tools~\cite{luu2016making,torres2018osiris,jiang2018contractfuzzer,he2019learning,wustholz2020harvey,
schneidewind2020ethor,kalra2018zeus,tsankov2018securify,grossman2017online,ferreira2020aegis,rodler2018sereum,
chen2020soda,Wu2020TimeTravelIT} mainly focus on the code vulnerability,
such as the re-entrancy and the integer overflow. 
However, to the best of our knowledge, they cannot be directly used
to detect \defi{} attacks caused by logic vulnerabilities due to the 
lack of the capability to recover and understand high-level \defi{}
semantics, e.g., a user trades a token pair \code{X} and \code{Y}
in a Decentralized EXchange (DEX).

Among DeFi apps, decentralized exchanges and lending apps
are the two most popular types of ones. According to the
statistics~\cite{dapp-rank-list}, there are five DEXes
and four lending apps in the top ten DeFi apps.
Automated Market Maker (AMM) is the most popular
type of DEX that provides the cryptocurrency exchange service.
Each pool in the AMM  maintains two or more types of
cryptocurrencies and leverages a  price mechanism
to decide the exchange rate. 
Besides, a user can borrow cryptocurrencies from a lending app
by depositing the collateral (e.g., the USDC token) into the app.
To determine how many cryptocurrencies that a user can borrow,
the lending app needs to retrieve the current price of the collateral,
e.g., from a pool in AMM.

We observe that with the popularity of DEXes and lending apps,
two types of new attacks are emerging, i.e.,
\textit{direct} and \textit{indirect} price manipulation attacks.
As the name suggests, the former
one means that an attacker \textit{directly} manipulates the token\footnote{In this paper,
we interchangeably use cryptocurrencies, assets, and tokens to denote Ether
and ERC20 tokens.}
price in a pool of an AMM. It is usually achieved through performing an unwanted
trade in the same pool by attacking the vulnerable DeFi app.
The latter one means that an attacker \textit{indirectly} manipulates
the token price of the vulnerable DeFi app (e.g., a lending app),
whose price mechanism depends on the \textit{real-time} status, e.g., the
quotation and reserves of a token, in an AMM.
An attacker can manipulate the status by making a
trade in the AMM. For instance, the attacker can raise up
the price of the collateral in the AMM pool that
provides the price information to the lending app
before borrowing loans. By doing so, the attacker can
borrow much more tokens than a legitimate borrower
can borrow with the same amount of collateral.

\smallskip \noindent \textbf{Our work}\tab
Our work aims to detect price manipulation attacks.
It requires us to analyze transactions between multiple
smart contracts and understand the high-level semantics of DeFi apps.
There are two challenges in this process.

\begin{itemize} [leftmargin=*]
\item{Challenge I: Complicated interactions.}\tab 
DeFi apps often consist of multiple smart contracts
that interact with each other. 
For example, {one attack transaction} of the Harvest Hack incident~\cite{Harvest-hack}
(explained in Section~\ref{subsec:show_example})
involves $1,316$ internal transactions. Fig.~\ref{fig:motivation1} shows
part of the internal transaction graph.
How to analyze the complicated interactions between smart contracts
is the first challenge.

\item{Challenge II: Semantic Gap.}\tab
There is a semantic gap between raw transactions that can be
observed on Ethereum and high-level \defi{} semantics
that are defined in DeFi apps. 
On Ethereum, we can only observe the field values of these (external or internal) transactions,
such as \textit{from}, \textit{to}, and \textit{input}.
However, we cannot get the high-level DeFi semantic such as
\textit{there exists an account that trades
$861.95$ USDC for $0.5$ Ether in the USDC-Ether pool using the Uniswap V2 protocol.}
This high-level DeFi semantic is critical to detect
price manipulation attacks since
the attack usually involves the trade of tokens.
\end{itemize}

Our work solves the two challenges in the following way.
First, our insight to solve the first challenge is that
though a \defi{-related} transaction
often triggers many internal transactions between smart contracts,
not all of them are useful for our analysis.
We can prune unnecessary ones.
Second, 
we propose a \textit{platform-independent} way to recover DeFi semantics.
We name it \textit{semantic lifting}
since it lifts low-level semantics in raw transactions
to  high-level DeFi semantics.

Specifically, we first collect raw Ethereum transactions and then construct
the cash flow tree (CFT in short). This tree reflects
the token transfer initialized by  an external (or internal) transaction and 
the corresponding account. We call the token
transfer as the basic DeFi action.
After that, our system recovers the high-level DeFi semantics
based on the token transfer behavior in CFT. In particular,
our system can automatically recognize three types of
advanced DeFi semantics, including the liquidity mining
(depositing tokens into an DeFi app),
liquidity cancel (withdrawing tokens from a DeFi app), and
the trade (of different tokens in an AMM pool). 
Finally, we detect the price manipulation attacks
using the detection patterns that are expressed with these
recovered DeFi semantics.

\smallskip \noindent \textbf{Evaluation}\tab
We have implemented a prototype system called \tool{}
and evaluated its effectiveness from two perspectives.
First, whether the proposed method can accurately recover
DeFi semantics? Second, whether our system can detect
real-world price manipulation attacks in the wild, including
zero-day ones?

To answer the first question, we compare the result of
recovered DeFi semantics with Etherscan~\cite{Etherscan}.
Our comparison shows that \tool{} can identify more
DeFi semantics than Etherscan, with a lower false negative
rate. What's more important, until the writing of this paper in
April 2021, Etherscan can identify none of the DeFi semantics
of vulnerable DeFi apps that are detected by \tool{}.
This clearly shows the value of our system to recover
DeFi semantics. Otherwise, none of the attacks can be detected.

To answer the second question, we performed a large-scale
\pricemani{} attack detection from $350,823,625$ transactions.
As a result, \tool{} detected $432$ real-world attacks in the wild.
We confirm that they belong to four known security incidents and five \textit{zero-day} ones, 
and they caused a total loss of around $39M$ USD. Note that,
one security incident can have hundreds of attacks, thus
the number of security incidents is less than the number of real attacks.
We reported our findings, and two CVEs have been assigned~\footnote{For
the anonymity purpose, we do not include the CVE numbers in the paper.}.
This result clearly shows that \tool{} can effectively detect
\pricemani{} attacks.

\smallskip \noindent \textbf{Attack Analysis}\tab
We further analyze the detected nine security incidents
to understand the root cause of the vulnerability,
the attack footprint, and the financial impact to
the vulnerable DeFi apps.
First, we find that vulnerabilities are either
due to insufficient access control of critical interfaces
in the smart contract or insecure price dependence.
Second, we find that attackers are becoming mature. They
take a \textit{clean attack strategy to make the trace of
attackers challenging}.
Third, security incidents have a direct impact on the
market value of vulnerable DeFi apps, which urges the need
to secure DeFi apps.  

\smallskip \noindent \textbf{Contributions}\tab
We summarize our main contributions as below.
\begin{itemize} [leftmargin=*]
    \item 
    We  first propose two \pricemani{} attacks
    and elaborate the challenges to detect them in the wild.
    Then we propose a method to recover DeFi semantics to
    facilitate the detection. \textit{To the best of our knowledge,
    our work is the first to systematically detect \pricemani{} attacks
    on DeFi apps.}

    \item We implement a prototype named \tool{} and
    apply it to perform a large-scale \pricemani{} attack 
    detection. It successfully detected $432$ real-world attacks.
    These attacks contribute to four known security incidents and five {zero-day} ones.
    After reporting our findings to vulnerable DeFi apps, two CVEs have
    been assigned. 
    
    \item We perform detailed analysis on confirmed security incidents
    to understand the root cause of the vulnerability, attack footprint, and financial impact to victim DeFi apps. Our analysis urges 
    a need to secure the DeFi ecosystem.

\end{itemize}

To engage the community, we will release \tool{} to the research community.

	\section{Background}
\label{sec:background}

\begin{figure}[t]
	\centering
	\ifx\arxiv\undefined
	\includegraphics[width=0.45\textwidth]{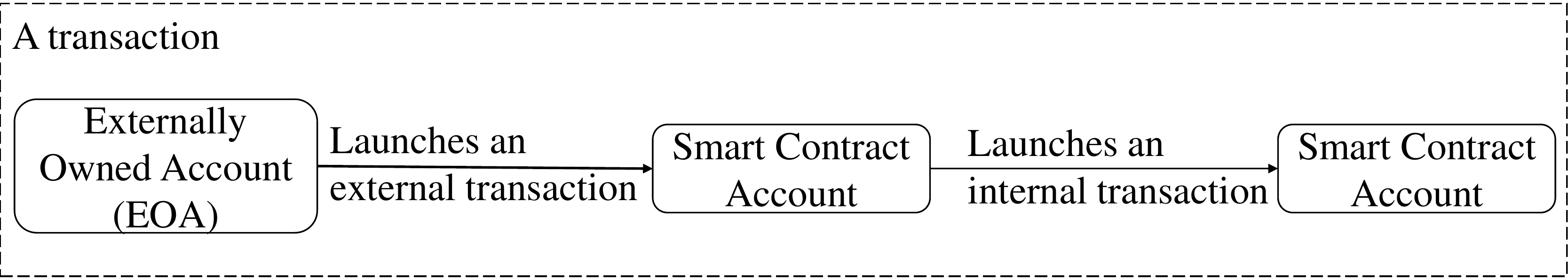}
	\else
	\includegraphics[width=0.6\textwidth]{pdfs/transactions.pdf}
	\fi
    \caption{A transaction that consists of an external transaction and an internal transaction.}
	\label{fig:transaction}
\end{figure}

This section presents necessary background information to better
understand our work.

\subsection{Ethereum Accounts and Transactions}

Ethereum has two types of accounts, i.e., External Owned Account (EOA) and smart contract account.
A transaction can be used to send Ether between accounts or to invoke
APIs in smart contracts. 
There are two types of transactions, i.e., external transaction and internal transaction. 
The external transaction is triggered by an EOA, while the internal transaction
is triggered by a smart contract. 
Specifically, an external transaction sent by
an EOA can invoke functions inside a smart contract, which will further create an internal transaction. 
The internal transaction can take a similar way to invoke a smart contract and trigger other internal transactions.
In the following of this paper, the usage of 
``transaction'' without specifying ``external'' or ``internal'' implies that we take an external transaction and the 
triggered internal transactions as a whole, as shown in Fig.~\ref{fig:transaction}.  

\subsection{Ethereum Cryptocurrencies }

There are two types of cryptocurrencies in Ethereum: Ether and ERC20 token. Ether is the native token in Ethereum, and an ERC20 token is
the third-party one that complies with the ERC20 standard. 
Every account, including EOA and the smart contract account,
can own Ether and ERC20 tokens. 
Though there are many types of ERC20 tokens, the following two are related to our work.

	\smallskip \noindent \textbf{Stablecoins}\tab
    Stablecoins are a class of cryptocurrencies that guarantee 
    price stability. Typically, stablecoins are 
    either directly/indirectly backed or intervened through different stabilization 
    mechanisms~\cite{moin2019classification,clark2019sok}. The popular
    stablecoins include USDC~\cite{usdc} or USDT~\cite{usdt}. 
    
    \smallskip \noindent \textbf{Liquidity Provider (LP) Tokens}\tab
    A \defi{} app may issue LP tokens to 
    motivate users to provide more liquidity (i.e., depositing cryptocurrencies
    into that \defi{} app). Users providing liquidity are called 
    liquidity providers. Liquidity providers can use the LP tokens as certificates to withdraw their deposits or 
    exchange for other cryptocurrencies in those decentralized exchanges
    (Section~\ref{subsec:defi}).
In this paper, we interchangeably use cryptocurrencies, assets, and tokens to
denote Ether and ERC20 tokens.

\subsection{Decentralized Finance (\defi{})}
\label{subsec:defi}
DeFi app usually consists of multiple
smart contracts to implement its functionality, running on the blockchain.
Some financial services, such as lending, exchange and portfolio management, have emigrated in the 
\defi{} ecosystem. As of this writing, more than 200 \defi{} apps have been   deployed  on Ethereum.

\smallskip \noindent \textbf{Decentralized EXchange (DEX)}\tab
DEX is an exchange where users can trade different tokens in
a decentralized way by interacting with the smart contract(s).
Compared with the traditional centralized exchange, DEX
possesses several distinct advantages, especially in privacy and capital 
management.

There are two modes in DEX, including List of Booking (LOB) and Automated Market Maker (AMM). DEX using the LOB mode 
maintains an off-chain order book to record users' bids and asks, namely,the matching of orders will be completed off-chain.
Alternatively, the AMM mode achieves a full decentralized exchange. The market maker may put two or more tokens 
into a liquidity pool with an equivalent or a self-defined weight.
The trade rate between cryptocurrencies in the pool 
will be calculated automatically based on the pricing mechanism.
AMM becomes the most popular mode in DEX due to its 
 flexible liquidity.

\smallskip \noindent \textbf{Lending}\tab
To borrow cryptocurrencies, borrowers are required to
over-collateralize other cryptocurrencies for covering 
the loan due to the pseudo-anonymity of Ethereum.

For example, in MakerDAO~\cite{maker}, borrowing 100 DAI 
requires tokens that are worth 150 Ether as collateral.
Moreover, once the collateral's value falls below a fixed threshold, it 
causes liquidation (the lending app will sell the collateral), as well as a designed penalty that will be applied to borrowers.

\smallskip \noindent \textbf{Flash Loan}\tab
Some DeFi apps~\cite{aave}~\cite{Uniswap-V2}~\cite{dydx} provide a type of non-collateral loan called 
flash loan. A valid flash loan "generously" lends users a considerable amount of capital without any collateral. 
The security of the loan is guaranteed since the user needs to borrow and
return the loan in a single transaction (with multiple following internal
transactions). Otherwise, the lending transaction will be reverted by
the loan provider. 

The flash loan gives everyone the ability to temporarily own a large number
of tokens. However, it can also be abused to launch attacks, 
and a number of such attacks have been observed in the wild~\cite{wang2020towards,qin2020attacking,cao2021flashot}.

\smallskip \noindent \textbf{Portfolio Management}\tab
As more and more \defi{} apps motivate clients to provide
liquidity, another kind of apps, known as portfolio 
management apps, debut to help users (liquidity providers)
to invest their cryptocurrencies. In particular,
they automatically find the apps that provide the highest Annual Percentage Yields (APY) and then 
invest clients' deposits to these apps.

	\section{Price Manipulation Attacks}
\label{sec:price_manipulation_attack}

In this work, we aim to detect two types of \pricemani{} attacks, including
the \textit{direct} and the \textit{indirect} ones. As the name suggests, the former
one means that an attacker \textit{directly} manipulates the token price
in a liquidity pool of an AMM. It is usually achieved through performing an unwanted
trade in the same pool by attacking the vulnerable DeFi app.
The latter one means that an attacker \textit{indirectly} manipulates
the token price of the vulnerable DeFi app, whose price mechanism
depends on the \textit{real-time} status, e.g., the
quotation and reserves of a token, in an AMM.
An attacker can manipulate the status by making a trade in the
AMM.

In the following, we first use the Uniswap V1
protocol~\cite{Uniswap-V1} as an example to introduce the
price mechanism of an AMM and then elaborate the two types of attacks.

\subsection{Price Mechanism of the Uniswap Protocol}

A liquidity pool of Uniswap makes up reserves of two
cryptocurrencies.
Users can trade tokens with the price determined by the
pricing mechanism/formula of Uniswap, as follows: 

\begin{equation}
    \label{formula:uniswap-pricing-mechanism}
    F(x) = \frac{997x * R_{Y}}{997x + 1000R_{X}}
\end{equation}

Specifically, $F(x)$ shows the amount of token \code{Y} that a user can trade
with the token \code{X} in an amount of $x$.
$R_{X}$ and $R_{Y}$ are the balances of the token \code{X} 
and token \code{Y} in the liquidity pool, respectively. 
Since the exchange rate between the token pair depends on the
pool's reserve balances (or reserves), attackers who
are capable of draining the 
pool can make a trade to inflate or reduce a token's price,
namely, manipulate the price deviating from the market price.
Such a behavior is referred to as \textit{\pricemani{}}.
DeFi apps that interact with a manipulated pool may suffer
from financial losses.

\subsection{Direct Price Manipulation Attack}
\label{subsec:direct_price_manipulation_attack}

Some DeFi apps have interfaces to trade tokens in AMMs.
However, if the interfaces are not properly protected,
an attacker can abuse these interfaces to trade tokens
on behalf of the vulnerable DeFi app, which affects the exchange
rate of a token pair. Then the attacker can make another trade of his or her
own token to gain profits. Since the token price is directly manipulated
by trading a token pair in an AMM, we call such an attack a {direct \pricemani{}  attack}.

\begin{figure}[t]
	\center
	\ifx\arxiv\undefined
	\includegraphics[width=0.45\textwidth]{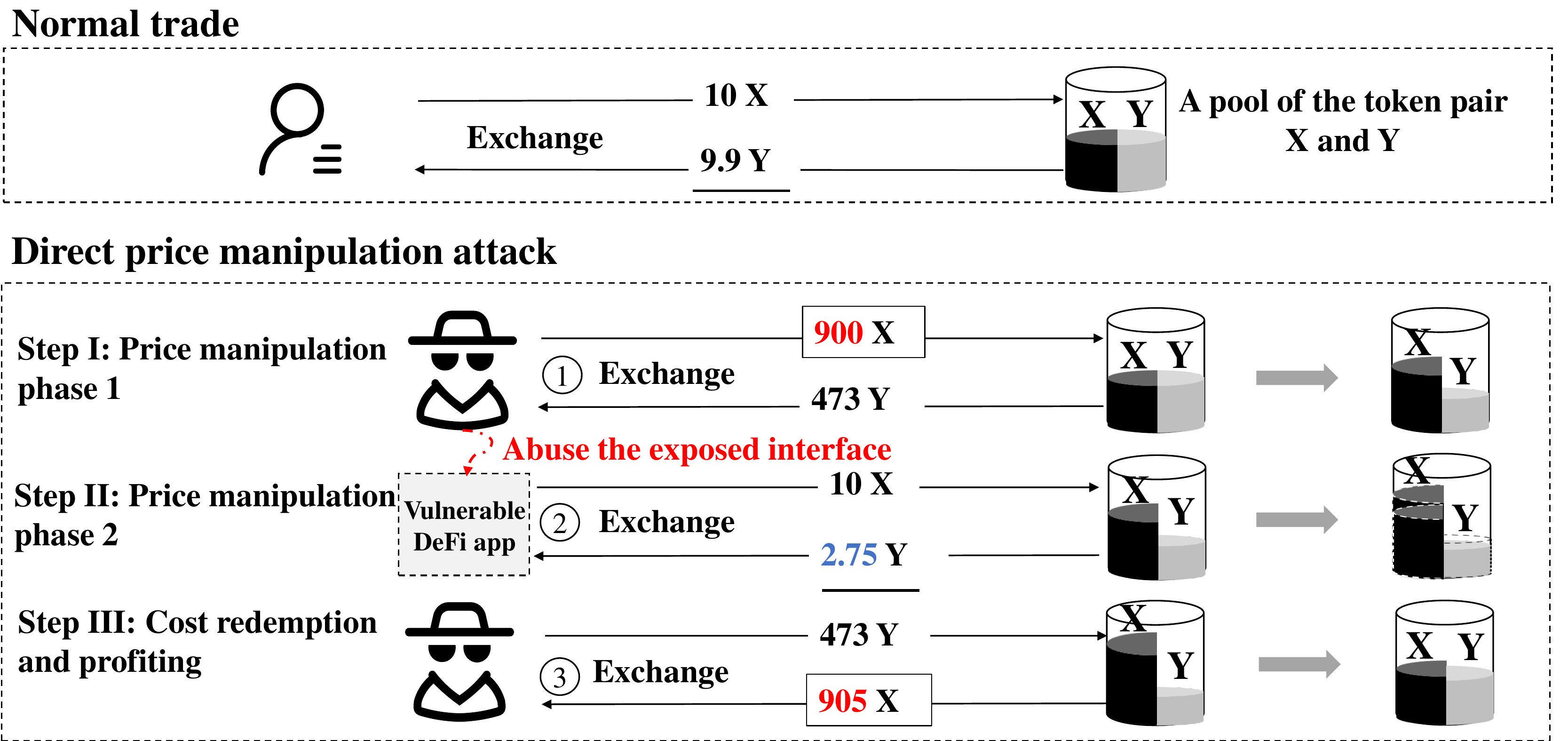}
	\else
	\includegraphics[width=0.6\textwidth]{pdfs/Direct_price_manipulation.pdf}
	\fi
	
    \caption{The \textit{direct} price manipulation attack. The initial balances
    	of the token pair are $1000$. The vulnerable DeFi app
    	exposes an interface
    	that can be exploited by the attacker to sell tokens
    	in the second step.}
	\label{fig:direct_price_manipulation}
\end{figure}

Fig.~\ref{fig:direct_price_manipulation} shows an example. Suppose
the pool has the same initial reserves ($1,000$) of a token pair 
\code{X} and \code{Y}. In a normal trade, a user can get $9.9$ \code{Y}
with $10$ \code{X},
according to the formula~(\ref{formula:uniswap-pricing-mechanism}).
An attacker can use the following three steps to perform the {direct}
\pricemani{} attack.

\smallskip
\noindent\textbf{Step I: Price manipulation phase one}\tab
    The attacker uses $900$ \code{X}, that occupies a large proportion of the pool to exchange for the token \code{Y},  which breaks the balance of the token pair and
    lifts the token \code{Y}'s price in the pool. 
    
    \smallskip
    \noindent{\textbf{Step II: Price manipulation phase two}}\tab
    The attacker invokes the public interface of the vulnerable DeFi app
    to sell $10$ \code{X}. However, the DeFi app only gets $2.75$ \code{Y} after consuming $10$ \code{X}.
    That's because the previous step reduces the price of the token \code{X}. 
    Besides, this trade further lifts the price of the token \code{Y} in the pool. 
    
    \smallskip
    \noindent{\textbf{Step III: Cost redemption and profiting}}\tab
    The attacker sells $473$ \code{Y} and gets $905$ \code{X} with a reverse exchange.
    That's because the token \code{Y}'s price has been raised in the second step. 
	By doing so, the attacker can get a profit of $5$ \code{X}.

Specifically, the first step inflates token \code{Y}'s price and reduces
token \code{X}'s price in the pool. This is an expected behavior
according to the formula~(\ref{formula:uniswap-pricing-mechanism}). 
However, the second step makes the vulnerable DeFi app
sell its token \code{X} and further raises the price of the token \code{Y}.
This is achieved by exploiting the exposed interface of
the vulnerable DeFi app.
As a result, the attacker can make a reverse exchange by selling
the token \code{Y} and get more \code{X} ($5$ in the example).
Compared with a 
normal trade ($10$ \code{X} with $9.9$ \code{Y}),
the victim DeFi app losses $7.15$ \code{Y} (i.e., $7.15 = 9.9 - 2.75$).

\subsection{Indirect Price Manipulation Attack}
\label{subsec:indirect_price_manipulation_attack}
Some DeFi apps need the token price for business purposes.
For instance, a lending app is required to 
calculate the collateral's price to decide how many tokens
a borrower is eligible to borrow. 
If the price mechanism of the lending app is manipulable,
a borrower might borrow more tokens
than the outstanding principal balance of the collateral 
(i.e., {undercollateralization~\cite{undercollateralized}}).

\begin{figure}[t]
	\center
	\ifx\arxiv\undefined
	\includegraphics[width=0.45\textwidth]{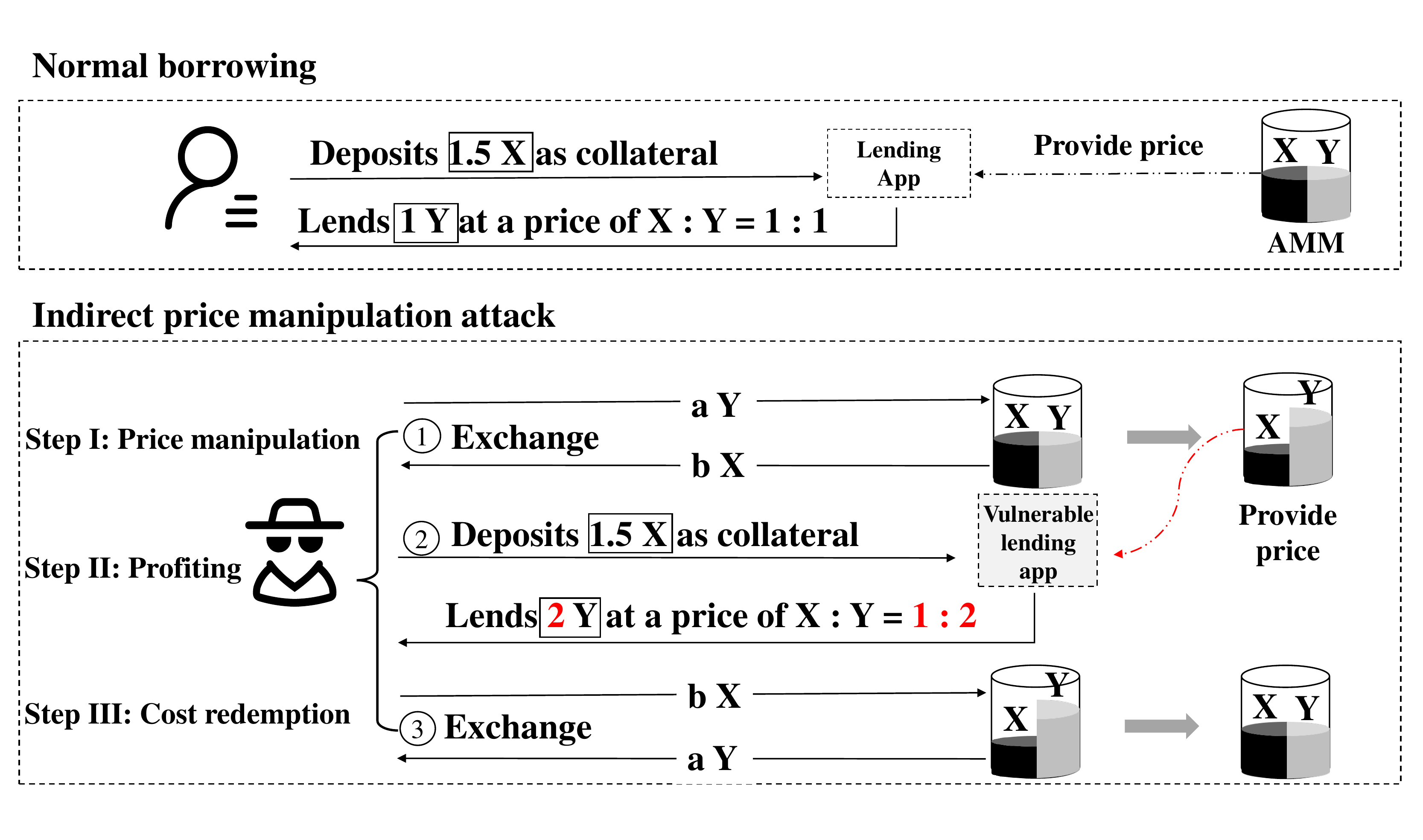}
	\else
	\includegraphics[width=0.6\textwidth]{pdfs/Indirect_price_manipulation.pdf}
	\fi
	
    \caption{The \textit{indirect} price manipulation attack. The vulnerable lending app uses the AMM's \textit{real-time} token price to determine the number of tokens that the borrower can borrow. 
		The {collateral ratio} of the lending app is $150\%$.}
	\label{fig:indirect_price_manipulation}
\end{figure}

Fig.~\ref{fig:indirect_price_manipulation} gives an example.
The lending app uses the real-time exchange rate of 
a token pair fetched from an AMM (through invoking the API exposed by the smart contract
of the  AMM) to determine the value of the collateral.
Suppose the initial exchange rate between the token
\code{X} and \code{Y} is 1:1.
In the normal borrowing scenario, a user deposits $1.5$ token \code{X} to
the lending app as the collateral and borrows $1$ token \code{Y},
since the {collateral ratio~\cite{collateralrate}}
of the lending app is $150\%$. 
An attacker is able to take the following
steps to launch the indirect price manipulation attack.

   \smallskip 
   \noindent{\textbf{Step I: Price manipulation}}\tab
	The attacker trades a tremendous amount of token \code{Y} for token \code{X},
	draining   out a large proportion of the token \code{X} 
	in the pool, thus  creating  an inflated price for the token \code{X}.
	Since the lending app's price mechanism depends on
	the AMM's \textit{real-time} quotation, 
	the token \code{X}'s price is also inflated in the lending app.

	\smallskip 
	\noindent{\textbf{Step II: Profiting}}\tab
	After manipulating the token \code{X}'s price,
	the attacker simply borrows the token \code{Y} using the token \code{X} as the collateral.
	In particular, he or she can borrow $2$ \code{Y} rather than
	$1$ \code{Y} with the same amount
	of the collateral ($1.5$ \code{X}) as in the normal borrowing scenario.

	\smallskip 
	\noindent{\textbf{Step III: Cost redemption}}\tab
	The attacker only needs to redeem the cost of \pricemani{} 
	by making a reverse exchange in the AMM pool.

The root cause of this attack is that the vulnerable lending app leverages
an AMM's \textit{real-time} quotation to decide the collateral's price.
As a result, the attacker can make a trade in the AMM's pool to 
affect the token price (Step I) and then borrow {an undercollateralized loan} from the lending app
(Step II). After that, the attacker makes a reverse trade to {redeem
	the cost} (Step III).

Note that the previous steps of these two attacks are simplified
for better illustration.
In a real attack, the attacker may leverage the 
flash loan~\cite{flashloan,wang2020towards,qin2020attacking,cao2021flashot}
to borrow a large number of tokens
that are required in the first step.

\begin{figure}[t]
	\center
	\ifx\arxiv\undefined
	\includegraphics[width=0.45\textwidth]{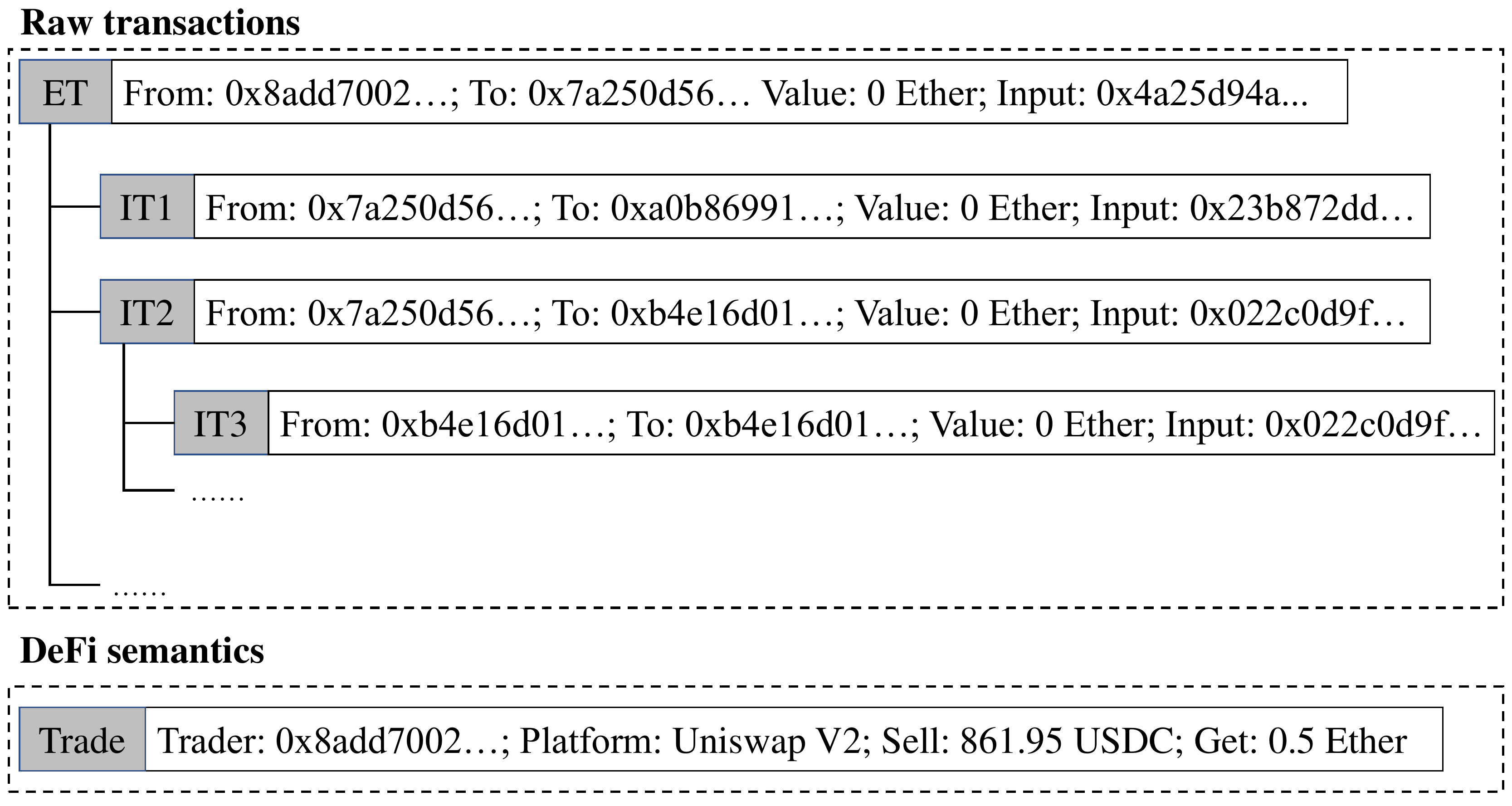}
	\else
	\includegraphics[width=0.6\textwidth]{pdfs/semantic_gap.pdf}
	\fi
    \caption{This is a token trade in the USDC-Ether pool of the Uniswap V2.
    	The trade triggers $11$ internal transactions. This figure shows
    	the gap between the raw transactions observed on Ethereum and
    	the high-level Defi semantic (the token trade).  }
    \label{fig:semantics_gap}
\end{figure}


	\section{Challenges and Solutions}

\begin{figure*}[t]
	\center	
    \includegraphics[width=0.8\textwidth]{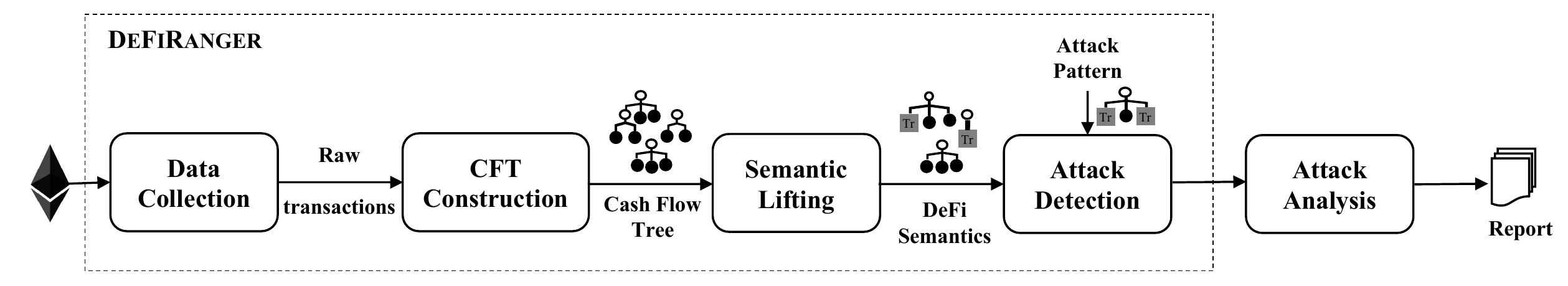}
    \caption{The overview of our work.}
	\label{fig:workflow}
\end{figure*}

In this section, we present two challenges in detecting price manipulation
attacks and then illustrate our methods to address these challenges.

\subsection{Challenges}
\label{subsec:challenge}
The price manipulation attacks originate from logic vulnerabilities of DeFi apps, 
which makes it non-trivial to perform the detection. 
Namely, detecting such attacks requires us to analyze transactions between multiple
smart contracts and understand the high-level semantics of DeFi apps.

\smallskip \noindent \textbf{Challenge I: Complicated interactions}\tab
DeFi apps tend to have complicated business logic
and consist of multiple smart contracts that interact with each other. 
For instance, the average number of smart contract invocations
triggered by each transaction before December $2019$ is $1.19$.
From December $2019$ to
April $2021$, the number becomes $2.40$, due to the DeFi boom.
Besides, \defi{-related} transactions between multiple smart contracts
are too complicated to be analyzed by existing tools~\cite{Etherscan,EthDecoder}. 
For example, {one attack transaction} of the Harvest Hack incident~\cite{Harvest-hack}
(explained in Section~\ref{subsec:show_example})
involves $1,316$ internal transactions. Fig.~\ref{fig:motivation1} shows
part of the internal transaction graph.
Obviously, it is challenging to analyze the complicated interactions between smart contracts.

\smallskip \noindent \textbf{Challenge II: Semantic gap}\tab
There exists a semantic gap between raw transactions that are observed on
Ethereum and \defi{} semantics that are defined in DeFi apps. 
As shown in Fig.~\ref{fig:semantics_gap}, the address \code{0x8add7002}
launches an external transaction that uses the Uniswap V2 protocol to 
exchange tokens. This external transaction further triggers $11$ internal transactions. 
On Ethereum, we can only observe the field values of these (external or internal) transactions,
such as \textit{from}, \textit{to}, and \textit{input}.
However, we cannot identify the high-level DeFi action that the address \code{0x8add7002} trades
$861.95$ USDC for $0.5$ Ether in the USDC-Ether pool using the Uniswap V2 protocol.
This high-level DeFi semantic is critical to detect
price manipulation attacks since
they usually involve the trade of tokens.

\subsection{Our Solutions}
 
Our work takes the following methods to address the two challenges.

\smallskip \noindent \textbf{Method I: Pruning unnecessary transactions}\tab
Our observation suggests that not all of the internal transactions triggered by a 
\defi{-related} transaction are helpful for our analysis.
As such, we can solve the first challenge by pruning those unnecessary ones.

For example, the transaction in Fig.~\ref{fig:semantics_gap} triggers $11$ internal transactions.
Only three of them are {closely} related to the token trade. The remaining ones
are used for auxiliary operations, such as finding the suitable pool and checking 
token balances.  
Furthermore, we observe that
the {token transfer}
(including token minting and burning)
is the primitive action in the \defi{} ecosystem. 
Therefore, we simplify the interaction in \defi{-related} transactions by removing
transactions that are unrelated to the 
{token transfer}.
We illustrate this process in
Section~\ref{subsec:CFT_Construction}.

\smallskip \noindent \textbf{Method II: Recovering high-level semantics}\tab
We recover high-level DeFi semantics from raw transactions to
solve the second challenge.
Note that, even though Etherscan~\cite{Etherscan} provides the recovered \defi{} semantics
on its website, the information is incomplete. 
The experimental result in Section~\ref{subsec:evaluation_action_indentification} shows that 
Etherscan~\cite{Etherscan} has a higher false negative {rate} in identifying
DeFi actions.
Our work takes a \textit{platform-independent} way to
recover DeFi actions. We name it \textit{semantic lifting}
since it lifts low-level semantics in raw transactions
to  high-level DeFi semantics.
We will illustrate the process in {Section~\ref{subsec:semantic_lifting}}.

\section{Methodology}
\label{subsec:workflow}

Fig.~\ref{fig:workflow} shows the overview of our work. In particular, we first
propose a system called \tool{} to detect price manipulation attacks. Then for
the detected attacks, we perform further analysis from multiple perspectives,
e.g., understanding the root cause of the vulnerability and the impact on the
vulnerable DeFi apps. We elaborate \tool{} {in this section} and attack analysis in Section~\ref{sec:forensic_analysis}, respectively.

\subsection{Overview}
\tool{} first collects raw Ethereum transactions (Section~\ref{subsec:datacollection}), and then
constructs the cash flow tree (Section~\ref{subsec:CFT_Construction}).
After that,
it recovers the DeFi semantics (Section~\ref{subsec:semantic_lifting})
and then uses predefined patterns
expressed with DeFi actions to
detect price manipulation attacks (Section~\ref{subsec:detection}).

\subsection{Data Collection}
\label{subsec:datacollection}
Though a few online services, including Google's BigQuery Ethereum service, provide interfaces to
query transactions, they are not enough to support the analysis of \tool{}. For instance, the internal states
of EVM when executing a smart contract are not available.
To serve the need, we deploy an Ethereum full node with a
modified Geth~\cite{Geth} client to collect the required data. 
{First}, we collect internal transactions' metadata, including \textit{from}, \textit{to}, \textit{input}, etc. 
{Second}, we collect the EVM depth when executing each
internal transaction. This helps us recover the
invocation order of different functions in multiple smart contracts.  
{Third}, we collect the execution order between internal
transactions and events. This can help
us to retrieve the sequence of token transfers.

\begin{figure*}[t]
	\center
	\includegraphics[width=0.8\textwidth]{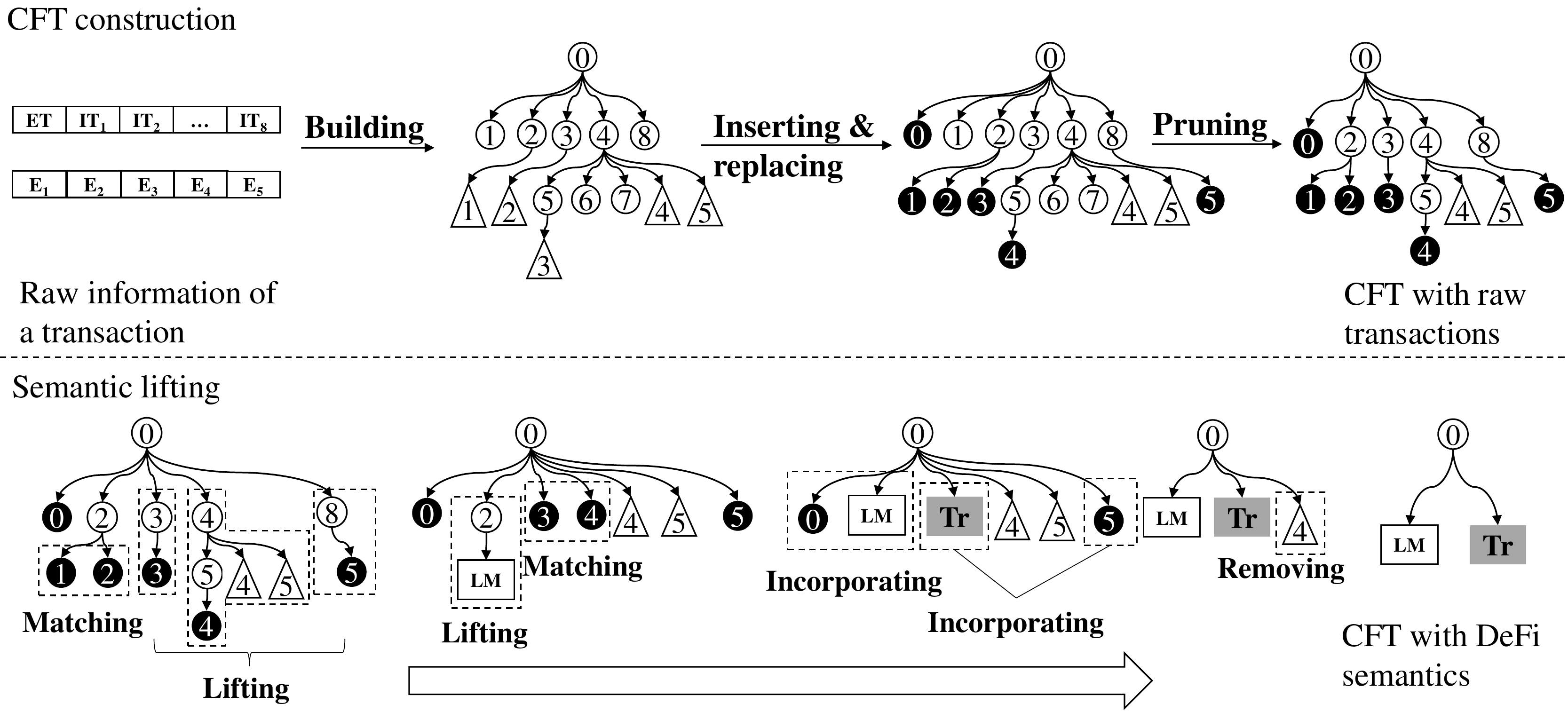}
	\caption{The whole process of the CFT construction and semantic lifting.
    ET: External transaction; IT: Internal transaction; E: Event; LM: \code{Liquidity mining}; Tr: \code{Trade};
    \protect\Bigcircle: Transaction node; \protect\Bigtriangle: Event node; \protect\Bigbullet: Transfer node.}
	\label{fig:construct_CFT}
\end{figure*}

\subsection{CFT Construction}
\label{subsec:CFT_Construction}
 
After collecting the Ethereum raw transactions, our system builds
the \textit{cash flow tree}~\cite{cashflow} ({CFT} for short).
The CFT is used to convert raw transactions to token transfer, which
lays a foundation to lift \defi{} semantics  (Section~\ref{subsec:semantic_lifting}).
For a better understanding, we use Fig.~\ref{fig:construct_CFT} as an example to 
illustrate the whole process.
There are three types of nodes inside the tree.

\begin{itemize} [leftmargin=*]
	\item{{Transaction node}}\tab Both external
	and internal transactions are included in the tree. The directed edge between two {(external or internal)} 
    transactions means that the child transaction is created by the parent transaction. For instance, the external
	transaction  \circled{0} creates an internal transaction \circled{1}. Thus, there is a directed
	edge from \circled{0} to \circled{1} in the graph. 
	
	\item{{Event node}}\tab Events are emitted by a smart contract during its execution for the logging purpose. A
	smart contract can selectively emit events. In the graph, events are child nodes of internal transactions since the event emitting is initialized by
	an internal transaction to execute a smart contract.
	
	\item{{Transfer node}}\tab Transfer node denotes the existence of Ether or {ERC20} token transfer that is initiated 
    by an external (or internal) transaction. 
\end{itemize}

\smallskip
\noindent \textbf{Building the tree structure}\tab
The first step is to build a tree from raw transactions, including internal and external ones. This
process is straightforward. We only need to append a child node to a parent
node if the child transaction is created from
the parent. 
For instance, five internal transactions (from \circled{1} to \circled{4} and \circled{8}) are created by one external
transaction, thus they are the child nodes of the parent node \circled{0}.
Note that, we also append the event node in the graph as the child of the transaction node that creates the event.
For instance, the event node $\trinum{1}$ is the child of node \circled{2} since it is created by this internal transaction. 
 
\smallskip
\noindent \textbf{Inserting and replacing with transfer nodes}\tab
With the tree structure built in the previous step, we need to append
transfer nodes in the tree. That's
because the token 
transfer is the basis for recovering DeFi semantics.
First, we append the Ether transfer node to the tree. We can easily retrieve such nodes
since the Ether transfer has a non-zero \textit{Value} field of a transaction. 
We insert the Ether transfer node as the first child of the corresponding transaction node, e.g.,
\encircled{1} in the figure.
Second, we leverage the standard event \code{Transfer} to locate the transfer of ERC20 tokens.
We replace \code{Transfer} event nodes with {ERC20} token transfer nodes, e.g., replacing  $\trinum{1}$ to \encircled{2}.
Note that, internal transaction \circled{2} creates two 
transfers, one is transferring Ether (\encircled{1}) and
the other is transferring the ERC20 token (\encircled{2}). 
 
\smallskip
\noindent \textbf{Pruning tree branches}\tab
Our system traverses the tree and prunes all branches
without a transfer node.
That's because such branches are not  helpful to lift semantics.
{This can greatly remove unnecessary interactions to address
the challenge I discussed in Section~\ref{subsec:challenge}. 
For instance, it removes $94.23\%$ ($1,241$) transaction nodes for a real-world attack in Section~\ref{subsec:show_example}}.
In Fig.~\ref{fig:construct_CFT}, transaction nodes \circled{1}, \circled{6} and \circled{7} are removed from the tree.

\begin{table*}[t]
	
	\caption{{The definition of \defi{} actions. $F_f$ means the value of the field $f$ inside an (external or 
	internal) transaction or an event that records a transfer of an ERC20 token.}}
	\small
	\resizebox{1.08\textwidth}{!}{
		\hskip-0.6in
		\begin{tabular}{|l|l|p{15mm}|p{50mm}|p{68mm}|p{60mm}|}
			
		\hline
		{\defi{} Action} & Symbol & Attributes & Condition & Assignment & Description \\
		\hline
		Basic actions & $Ba$ & & & & Basic actions include the following three types of \code{transfer}: $T$, $T_m$, and $T_b$. \\
		\hline
		{Transfer(normal)} & $T$ & $spender$, \newline $recipient$, \newline $amount$, \newline $asset$ 
			& $F_{spender} !=  F_{recipient}$ \newline 
			and $F_{spender}!=$ zeroaddress \newline 
			and $F_{recipient}!=$ zeroaddress \newline 
			and $F_{amount}>$ 0 
			& $spender=F_{spender}$; \newline
			$recipient=F_{recipient}$; \newline
			$asset=F_{asset}$; \newline
			$amount=F_{amount}$
			& An account ($spender$) transfers a certain amount ($amount$) of certain asset ($asset$) to another account ($recipient$).  \\ 
		\hline			
		{Transfer(minting)} & $T_m $ & $spender$, \newline $recipient$, \newline $amount$, \newline $asset$ 
			& $F_{spender}==$ zeroaddress; \newline 
			and $F_{amount}>$ 0; 
			& $spender=F_{spender}$; \newline
			$recipient=F_{recipient}$; \newline
			$asset=F_{asset}$; \newline
			$amount=F_{amount}$
			& An ERC20 token contract mints a certain amount ($amount$) of certain asset ($asset$) to an account ($recipient$). \\
		 \hline			
		{Transfer(burning)} & $T_b $ & $spender$, \newline $recipient$, \newline $amount$, \newline $asset$ 
			& $F_{recipient}==$ zeroaddress   \newline 
			and $F_{amount}>$ 0 
			& $spender=F_{spender}$; \newline
			$recipient=F_{recipient}$; \newline
			$asset=F_{asset}$; \newline
			$amount=F_{amount}$
			& An account ($spender$) burns a certain amount ($amount$) of certain asset ($asset$). \\
		\hline
		Advanced  actions & $Aa$ & & & & Advanced actions include the below three ones: $LM$, $LC$, and $Tr$. \\
		\hline
		{Liquidity Mining} & $LM$ & $operator$, \newline $recipient$, \newline $pool$, \newline $asset\_in$, \newline $asset\_out$, \newline $amount\_in$, \newline $amount\_out$
			& Exists $T$ and $T_m$ \newline 
			and $T.asset!=T_m.asset$ 
			& $operator=T.spender;$ \newline
			$recipient=T_m.recipient;$ \newline
			$pool=T.recipient;$ \newline
			$asset\_in=T.asset;$ \newline
			$asset\_out=T_m.asset;$ \newline
			$amount\_in=T.amount;$ \newline
			$amount\_out=T_m.amount$
			& An account ($operator$) of a liquidity provider deposits certain amount ($amount\_in$) of certain asset (${asset\_in}$) to provide liquidity in a liquidity pool ($pool$) of a \defi{} app, 
			which then mints certain amount ($amount\_out$) of its LP token ($asset\_out$) to an account ($recipient$) that is specified by the liquidity provider.  \\
		\hline
		{Liquidity Cancel} & $LC$ & $operator$, \newline $recipient$, \newline $pool$, \newline $asset\_in$, \newline $asset\_out$, \newline $amount\_in$, \newline $amount\_out$
			& Exists $T$ and $T_b$ \newline 
			and $T.asset!=T_b.asset$ 
			& $operator=T_b.spender;$ \newline
			$recipient=T.recipient;$ \newline
			$pool=T.spender;$ \newline
			$asset\_in=T_b.asset;$ \newline
			$asset\_out=T.asset;$ \newline
			$amount\_in=T_b.amount;$ \newline
			$amount\_out=T.amount$
			& An account ($operator$) of a liquidity provider burns a certain amount ($amount\_in$) of certain \defi{} app's LP token ($asset\_in$) to redeem deposits, 
			a pool ($pool$) of the app then transfers a certain amount ($amount\_out$) of certain asset ($asset\_out$) to an account ($recipient$) that is specified by the liquidity provider. \\
		\hline 
		{Trade} & $Tr$ & $operator$, \newline $recipient$, \newline $pool$, \newline $asset\_in$, \newline $asset\_out$, \newline $amount\_in$, \newline $amount\_out$
			& (Exists two \textit{transfers(normal)} \newline
			 $T1$ and $T2$ \newline
			and $T1.asset!=T2.asset$ \newline
			and $T1.recipient==T2.spender$) \newline
			or (exists $LM$ and $LC$ \newline
			and $LM.recipient$$==$$LC.operator$ \newline
			and $LM.asset\_in!=LC.asset\_out$ \newline
			and $LM.asset\_out!=LC.asset\_in$) 
			& $operator=T1.spender$ or ($LM.operator$); \newline
			$recipient=T2.recipient$ or ($LC.recipient$); \newline
			$pool=T1.recipient$ or ($LM.recipient$); \newline
			$asset\_in=T1.asset$ or ($LM.asset\_in$); \newline
			$asset\_out=T2.asset$ or ($LC.asset\_out$); \newline
			$amount\_in=T1.amount$ or ($LM.amount\_in$); \newline
			$amount\_out=T2.amount$ or ($LC.amount\_out$)
			& An account ($operator$) of a trader sells a certain amount ($amount\_in$) of certain asset ($asset\_in$) 
			for a certain amount ($amount\_out$) of certain asset ($asset\_out$) in a liquidity pool ($pool$) of an AMM, 
			and the trader specifies an account ($recipient$) as recipient. \\ 
		 						    
		 \hline		
 
		\end{tabular}
	}
	\label{tab:defiaction}
	\end{table*}
\subsection{Semantic Lifting}
\label{subsec:semantic_lifting}
Based on the CFT, we can further recover DeFi semantics that are
critical to detect price
manipulation attacks. In this work, we first define DeFi actions and then
introduce the algorithm to lift the DeFi semantics from the CFT.

\subsubsection{DeFi actions}
In this work,
we focus on the following four DeFi actions. More actions could be recovered if they are necessary
in the future.
Table~\ref{tab:defiaction} shows the definition.

\smallskip
\noindent\textbf{{Transfer}}\tab
    Transfer means a token  (\textit{asset})  is transferred 
    from one address (\textit{spender}) to another (\textit{recipient}). 
    Besides, ERC20 token standard~\cite{erc20tokenstandard} defines
    that when the \textit{spender} field is set to the zero address~\footnote{Zero address: 0x0000000000000000000000000000000000000000}, it means
    a token mining, i.e., directly depositing tokens into the
    address (\textit{recipient}). Similarly, if the \textit{recipient}
    is set to the zero address, then the tokens are burning.
    They are denoted as $T$, $T_m$ and $T_b$, respectively
    in Table~\ref{tab:defiaction}.

\smallskip
\noindent\textbf{{Liquidity Mining} and {Liquidity Cancel}}\tab 
    To get more liquidity, \defi{} apps issue LP tokens
    to create incentives for
    users to provide liquidity (deposit cryptocurrencies), known as liquidity mining ($LM$).
    Besides, liquidity providers can redeem cryptocurrencies with the LP tokens as certificates, which is
    named liquidity cancel ($LC$).
    Accordingly, the {liquidity mining} consists of two components, i.e., depositing users' liquidity
    ($T$) and minting DeFi app's LP token ($T_m$). The
    liquidity cancel consists of
    burning LP tokens ($T_b$) and redeeming the deposited liquidity ($T$).

\smallskip
\noindent\textbf{{Trade}}\tab
    In the normal situation, a {trade} consists of two {transfers}. Therefore, we combine a pair of 
    {transfers}, that transfer different assets and have a pivot address ($T1.recipient$ or $T2.spender$) as a 
    {trade}. We further assign the pivot address to the $pool$ attribute of the {trade}.
    That's because traders
    exchange their tokens through the liquidity pool (as the intermediary).

\begin{table*}[t]
	\centering
	\caption{Patterns of DeFi actions to detect \pricemani{} attacks.}
	\small
	\resizebox{1.0\linewidth}{!}{
		\hspace{-0.3in}
    \begin{tabular}{|p{20mm}|l|p{60mm}|p{55mm}|}
    
    \hline
        &  Combination of \defi{} actions & Detection rules & Description \\
        \hline
        {Direct price \newline manipulation \newline attack} & $Tr1 \rightarrow^{\rm a} Ba$ (or $Tr3$) $\rightarrow Tr2$ 
            &  Exists a sequence of \defi{} actions: $Tr1$, $Ba$ (or $Tr3$), and $Tr2$ \newline
                and $Tr1.operator==Tr1.recipient$ \newline
                and $Tr2.operator==Tr2.recipient$ \newline
                and $Tr1.operator==Tr2.operator$ \newline
                and $Tr1.pool==Tr2.pool$ \newline
                and $Tr1.asset\_in==Tr2.asset\_out$ \newline
                and $Tr1.asset\_out==Tr2.asset\_in$ \newline
                and $Tr1.amount\_out==Tr2.amount\_in$ \newline
                and $Tr1.amount\_in<Tr2.amount\_out$ \newline
                and (($Ba.spender!=Tr1.operator$ \newline
                and $Ba.recipient==Tr1.pool$ \newline
                and $Ba.asset!=Tr1.asset\_out$) \newline
                or ($Tr3.operator!=Tr1.operator$ \newline
                and $Tr3.pool==Tr1.pool$ \newline
                and $Tr3.asset\_out==Tr1.asset\_out$))
          & An attacker ($Tr1.operator$) inflates an asset's ($Tr1.asset\_out$) price in a liquidity pool ($Tr1.pool$) of an AMM via a {trade} ($Tr1$)
          and then abuses the exposed interface of a vulnerable app ($Ba.spender$ or $Tr3.operator$) to further inflate the asset's ($Tr1.asset\_out$) price in the pool on behalf of the vulnerable app.
          Finally, the attacker launches a reverse {trade} ($Tr2$) in the same pool ($Tr1.pool$) to gain profits. 
          Above three steps are consistent to the three steps in Section~\ref{subsec:direct_price_manipulation_attack}, 
          but we extend the middle step to {transfer} and {trade} for detecting more attacks.\\
        \hline
        {Indirect price \newline manipulation attack} & $Tr1 \rightarrow Ba$ (or $Aa$) $\rightarrow Tr2$
            & Exists a sequence of \defi{} actions: $Tr1$, $Ba$ (or $Aa$), and $Tr2$ \newline
                and $Tr1.operator==Tr1.recipient$ \newline
                and $Tr2.operator==Tr2.recipient$ \newline
                and $Tr1.operator==Tr2.operator$ \newline
                and $Tr1.pool==Tr2.pool$ \newline
                and $Tr1.asset\_in==Tr2.asset\_out$ \newline
                and $Tr1.asset\_out==Tr2.asset\_in$ \newline
                and $Tr1.amount\_out==Tr2.amount\_in$ \newline
                and $Tr1.amount\_in==^{\rm b}Tr2.amount\_out$ \newline
                and ($Ba.recipient==Tr1.operator$ \newline
                or ($Aa.pool!=Tr1.pool$ \newline
                and $Aa.recipient==Tr1.operator$))
            & An attacker ($Tr1.operator$) leverages a trade ($Tr1$) to indirectly manipulate the price in a vulnerable app ($Ba.spender$ or $Aa.pool$) 
            and then profits from the vulnerable DeFi app via certain \defi{} action ($Ba$ or $Aa$).
            Finally, the attacker launches a reverse {trade} ($Tr2$) to redeem costs. 
            Similarly, above three steps are consistent to the three steps in Section~\ref{subsec:indirect_price_manipulation_attack},
            but we extend the middle step to any \defi{} action that can profit the attacker for detecting more attacks. \\
        \hline
        Arbitrage & $Tr1 \rightarrow Tr2 \rightarrow Tr3 \rightarrow ... \rightarrow Trn$ 
            & Exists a sequence of \code{trades}: $Tr1$, $Tr2$, ..., and $Trn$ \newline
                and $Tr1.asset\_in==Trn.asset\_out$ \newline
                and $Tr1.asset\_out==Tr2.asset\_in$ \newline
                and $Tr2.asset\_out==Tr3.asset\_in$ \newline
                and ...... \newline
                and $Trn-1.asset\_out==Trn.asset\_in$ 
            & An arbitrager leverage a sequence of \code{trades} ($Tr1$, $Tr2$, ..., and $Trn$) to profit from the 
            price differences crossing multiple exchanges. \\
    \hline
    \end{tabular}
	}
\begin{tablenotes}
    \item $\rm a$: $\rightarrow$ represents the execution order; $\rm b$: for better illustration, we do not consider the fee charged by AMM.
\end{tablenotes}

	\label{tab:detection_rules}
	\end{table*}

\subsubsection{Lifting algorithm}
With the well-defined \defi{} actions, we then propose
{an} algorithm to lift CFT's semantics. 
Algorithm~\ref{alg:lifting}  post-order traverses the
CFT (\code{LiftLeaves}). For each non-leaf node, 
the function repeatedly merges its child nodes
in pairs until they cannot be merged (line $3$-$11$).

\begin{algorithm}[t]
    \caption{Lift the semantics of CFT}
    \label{alg:lifting}
    \SetKwInput{KwInput}{Input}                
    \SetKwInput{KwOutput}{Output}              
    \DontPrintSemicolon

    \KwInput{A CFT}
    \KwOutput{A CFT with lifted semantics}

    \SetKwFunction{FMain}{Main}
    \SetKwFunction{FLiftLeaves}{LiftLeaves}
    \SetKwFunction{FMergeSubtree}{MergeSubtree}

    \SetKwProg{Fn}{Function}{:}{}
    \Fn{\FMergeSubtree{root}}{
        newChildren = []\;
        \For{child $\in$ root.children} {
            \For{len(newChildren) $>$ 0} {
                tailChild = newChildren.pop()\;
                mergedChild, isMerge = MergeLeaves(tailChild, child)\;
                \If{isMerge} 
                {
                    child = mergedChild\;
                }
                \Else
                {
                    break \;
                }
            }
            newChildren.append(child)\;
        }
        root.children = newChildren\;
        \If{len(root.children) == $1$}
        {
            \KwRet root.children\;
        }
        \KwRet []\{root\}\;
    }

    \SetKwProg{Fn}{Function}{:}{}
    \Fn{\FLiftLeaves{root}}{
        \If{root has no child}
        {
            \KwRet []\{root\}\;
        }
        \Else
        {
            newChildren = []\;
            \For{child $\in$ root.children} {
                newChildren.extend(LiftLeaves(child))\;
            }
            root.children = newChildren\;
            \KwRet MergeSubtree(root)\;
        }
    }

    \SetKwProg{Fn}{Function}{:}{}
    \Fn{\FMain{CFT}}{
        LiftLeaves(CFT.root)\;
        RemoveRedundantEvents(CFT.root)\;
    }

    \;
\end{algorithm}

The algorithm follows two guidelines to merge the nodes
\code{MergeLeaves} (line $6$). 
First, if two adjacent leaves match the conditions of a 
\defi{} advanced action defined in Table~\ref{tab:defiaction}, they will be 
merged as the \defi{} advanced action. This is called
\textit{matching} in this paper.
{
For instance, transfer nodes \encircled{3} and \encircled{4} are merged into a $Tr$ node. }

{
Second,
if there exists a transfer chain in the two adjacent leaves, one will incorporate another one.
For example, two transfers are 
``A transfers $x$ Ether to B'', and ``B transfers $x$ Ether to C'', the former one will incorporate the latter one. 
This is called \textit{incorporating} in this paper.
For instance, transfer node \encircled{0} and transfer node \encircled{1} in the {liquidity mining} node 
constitute a transfer chain, then the $LM$ node incorporates transfer node \encircled{0}.}

Besides, if a subtree consists of only one leaf,
its parent node will be removed, 
and it will be lifted (line $13$-$14$) to form more adjacent leaves. 
In the end, it will remove redundant events on the CFT (line $27$).
For a better understanding, we name them as lifting and removing in Fig.\ref{fig:construct_CFT}.
{For instance, the subtree of the transaction node
	\circled{2} consists of only one leaf (the $LM$ node), 
then our algorithm removes the transaction node \circled{2} and lifts its subtree.
Besides, it also removes the redundant event node $\trinum{4}$.}

\subsection{Attack Detection}
\label{subsec:detection}
After recovering \defi{} semantics,
We detect \pricemani{} attacks by matching recovered 
semantics with attack patterns (or rules).
We then manually confirm the detected transactions. 
Table~\ref{tab:detection_rules} 
shows the rules.

\smallskip \noindent \textbf{Detecting direct \pricemani{} attacks}\tab
The rule follows the attack flow discussed in 
Section~\ref{subsec:direct_price_manipulation_attack}. 
Specifically, the first step and the third step are identified
by a pair of reverse {trade} ($Tr1$ and $Tr2$). 
The second step is a {trade} ($Tr3$) or a {transfer} 
($Ba$ including $T$, $T_m$, and $T_b$).
That's because a {transfer} can also influence the reserves of
a liquidity pool of an AMM. 

\smallskip \noindent \textbf{Detecting indirect \pricemani{} attacks}\tab
Similarly, the rule follows 
the flow discussed in 
Section~\ref{subsec:indirect_price_manipulation_attack}.
Particularly, we detect step I  and step II  by
identifying a pair of reverse 
 {trades} ($Tr1$ and $Tr2$). We extend the step II (profiting) from borrowing (that is often a  {transfer})
to any action ($Ba$ or $Aa$) that can profit the attacker.

\begin{figure}[t]
	\center
	\ifx\arxiv\undefined
	\includegraphics[width=0.35\textwidth]{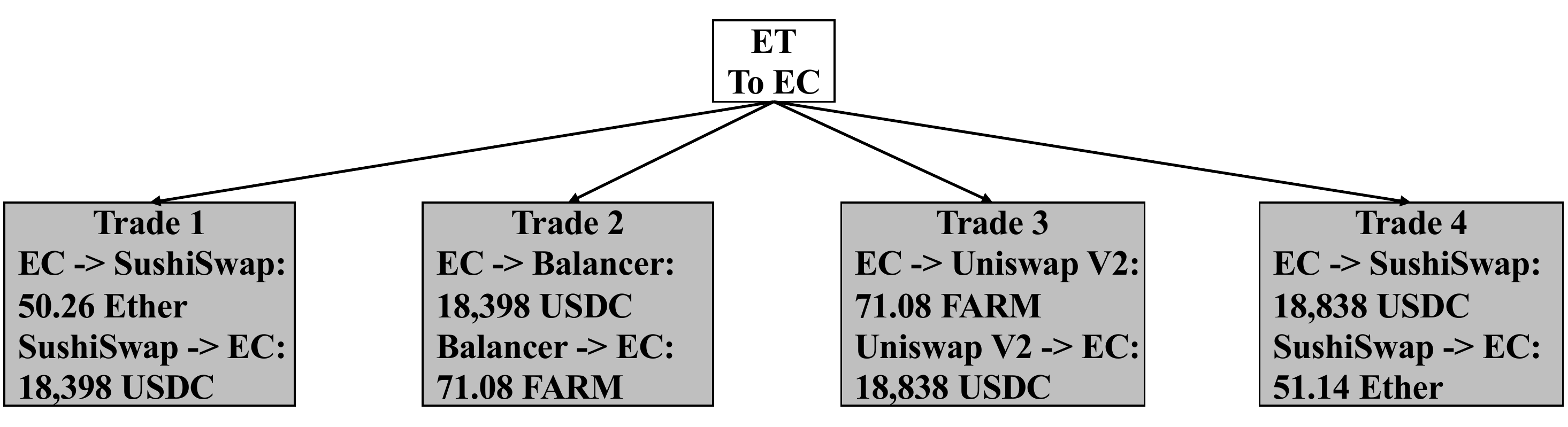}
	\else
	\includegraphics[width=0.6\textwidth]{pdfs/complex_arbitrage.pdf}
	\fi
	
    \caption{An arbitrage transaction. 
    ET: External Transaction, EC: Entry Contract.}
    \label{fig:complex_arbitrage}
\end{figure}

Note that, the reverse  {trade} pairs widely appear in
arbitrages~\cite{zhou2021just}, which will cause false positives.
We need to identify arbitrages and then remove them from the result.
Our observation is that arbitrage bots often only hold one type of asset.
They first purchase the asset used for arbitrage  
and then sell it back for the original asset afterwards.
As shown in Fig.~\ref{fig:complex_arbitrage}, the arbitrage bot (EC) only holds Ether.
It purchases the asset (USDC) used for arbitrage in {trade $1$} and gets Ether back in {trade $4$}. 
The {trade $1$} and {trade $4$} constitute a pair of reverse {trades}. 
In addition, the {trade $2$} and {trade $3$}, that capture the arbitrage opportunity also constitute a pair 
of reverse {trades}. 
The rules to detect arbitrages  are presented in Table~\ref{tab:detection_rules}.

\begin{figure*}[t]
    \includegraphics[width=1.0\textwidth]{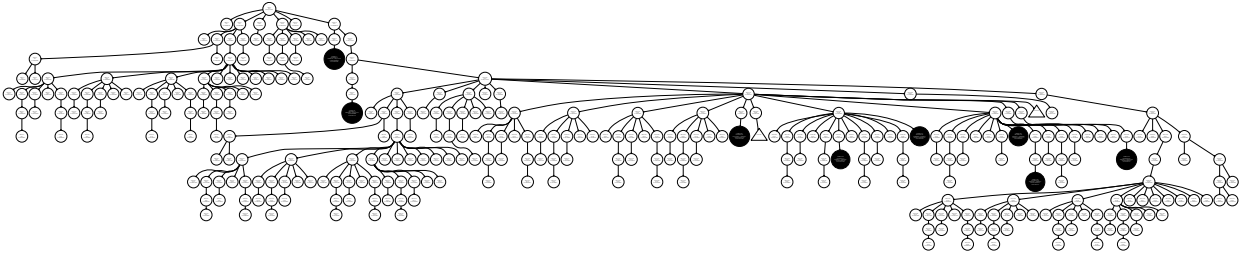}
    \caption{
    This CFT tree that describes the attack transaction in 
    Harvest Hack~\cite{Harvest-hack}. 
    The tree has $1,317$ transaction nodes, $58$ transfer nodes and $21$ event nodes. 
    Due to the space limit, we only draw a quarter of the tree.
    \protect\Bigcircle: Transaction node; \protect\Bigtriangle: Event node; \protect\Bigbullet: Transfer node.
    }
    \label{fig:motivation1}
\end{figure*}

\begin{figure}[t]
	\center
	\ifx\arxiv\undefined
	 \includegraphics[width=0.4\textwidth]{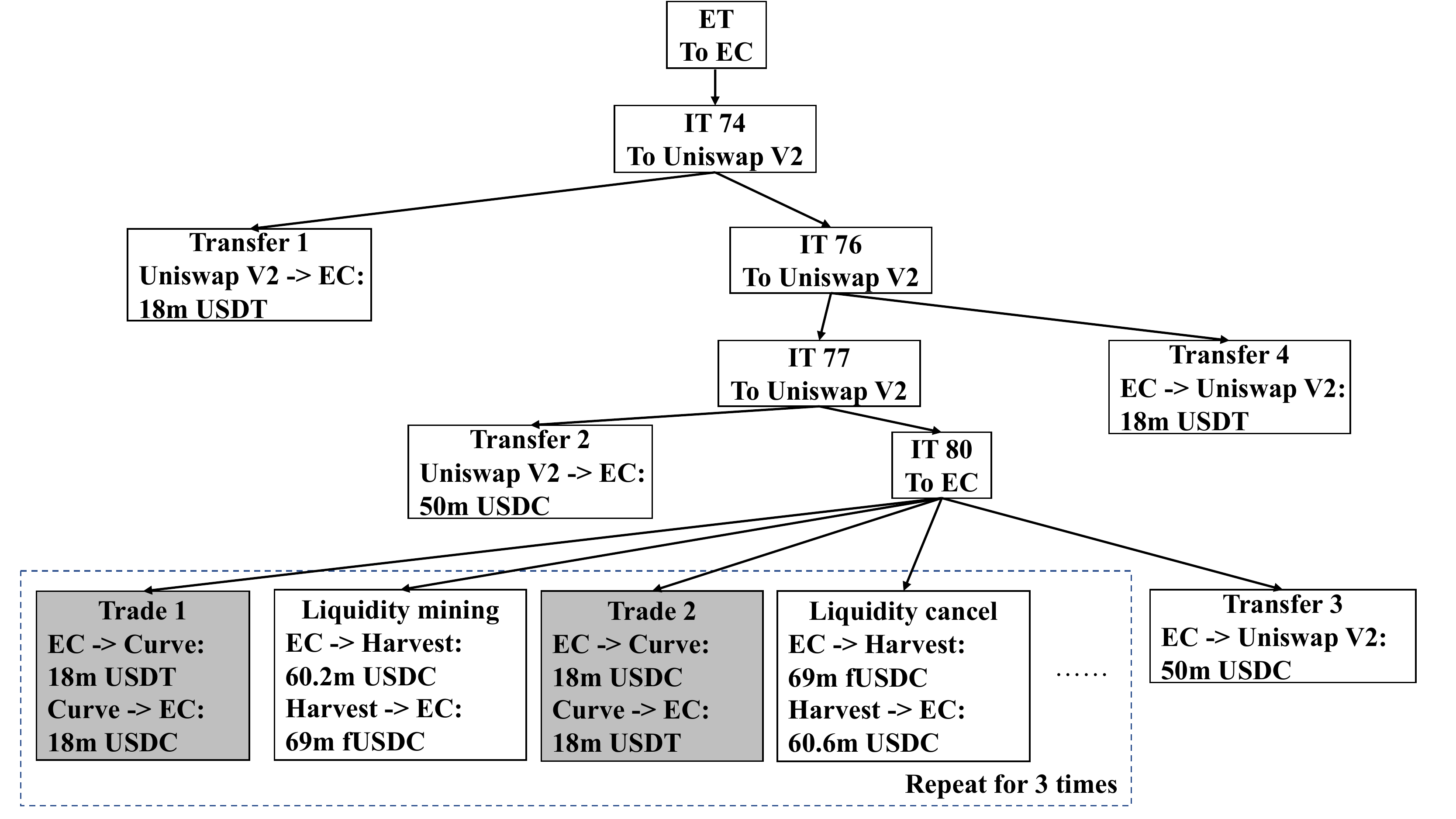}
	\else
	 \includegraphics[width=0.6\textwidth]{pdfs/motivation_example2_v2.pdf}
	\fi

    \caption{The CFT with \defi{} semantics. The figure is converted from the tree in Fig.~\ref{fig:motivation1} via two steps: 
    pruning branches (for challenge I) and semantic lifting (for challenge II).
    We change the shape of nodes to a box to facilitate the showing of node metadata.
    ET: External Transaction; IT: Internal Transaction; EC: Entry Contract.}
    \label{fig:motivation2}
\end{figure}

\subsection{A Real Example}
\label{subsec:show_example}

We use a real attack transaction~\footnote{\code{0xb460b70f11a93364fecf1f3c3ec49f053aecd2d6d9912c01-}
	\code{2170aa7a0de2d526}} in Harvest Hack~\cite{Harvest-hack} to show the entire process.
	The security incident made Harvest lose almost $24M$ USD.

\tool{} transforms raw transactions into a CFT
with \defi{} semantics via two steps, i.e.,
CFT construction and semantic lifting. 
During this process, it prunes branches in CFT
to address challenge I  (Section~\ref{subsec:challenge})  and uses the semantic lifting
to address challenge II (Section~\ref{subsec:challenge}).
The comparison of the transaction graph in Fig.~\ref{fig:motivation1}
and the CFT with DeFi semantics in Fig.~\ref{fig:motivation2}
shows the effectiveness of our system to solve the challenges.

In the attack detection process, 
\tool{} identifies {trade 1} and {trade 2} in Fig~\ref{fig:motivation2}
as a pair of reverse {trades}
($Tr1$ and $Tr2$), 
the entry contract (EC) as the attacker ($Tr1.operator$), and Curve as the manipulated pool ($Tr1.pool$). 
Furthermore, in the {liquidity mining}, the attacker
deposits $60.2m$ USDC to provide liquidity to the third-party 
app.
{Harvest then mints $69m$ fUSDC to the attacker as certificates ($LM.pool!=Tr1.pool$, $LM.recipient==Tr1.operator$)}.
Since the detection rule is satisfied,
\tool{} labels it as a suspicious {indirect} \pricemani{} attack.

	\section{Evaluation}
\label{sec:evaluation}

In this section, we evaluate \tool{} from two perspectives:
\defi{} action identification and \pricemani{} attack detection, i.e., whether it can accurately identify DeFi actions, and whether
it can detect real \pricemani{} attacks.

\subsection{\defi{} Action Identification}
\label{subsec:evaluation_action_indentification}

To the best of our knowledge, there is no ground truth for labeled DeFi actions in Ethereum.
Etherscan~\cite{Etherscan} is the only public online service that identifies \defi{} actions for
each transaction. To evaluate the effectiveness of our method, we compare our result with
that of Etherscan.
In particular, we crawled $183,363$ transactions~\footnote{All transactions were downloaded on January 30, 2021. Etherscan has a speed limit
and anti-crawling mechanism. We did not crawl more transactions to honor these mechanisms. } from Etherscan.
Then we apply our system on the downloaded transactions. Finally, we compare the identified
actions with the labels reported by Etherscan.

\begin{figure}[t]
	\center
	\includegraphics[width=0.48\textwidth]{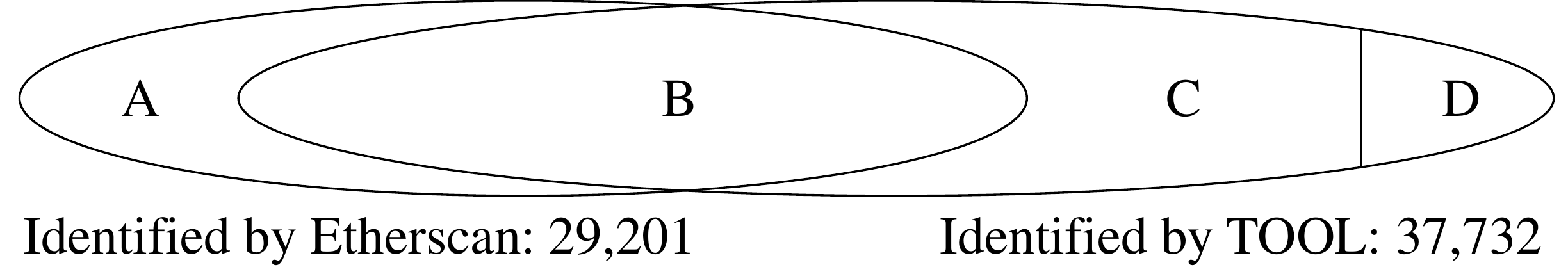}
    \caption{Distribution of identified \defi{} actions.
    A: Actions that Etherscan identifies but \tool{} misses; B: Actions that Etherscan and \tool{} both identifies;
    C: Actions that Etherscan misses but \tool{} correctly identifies; D: Actions that Etherscan misses but \tool{} incorrectly identifies.}
	\label{fig:compaison}
\end{figure}

For the same $183,363$ transactions, Etherscan and \tool{}
identify $29,201$ and 
$37,732$ \defi{} actions, as shown in the left and right oval in Fig.~\ref{fig:compaison},
respectively.
The number of actions that are identified by both Etherscan and \tool{} is 
$25,431$ (B), while the number of actions that are identified by Etherscan  but
are missed by \tool{} is $3,770$ (A).
Besides, \tool{} identifies extra $12,301$ (C+D) DeFi actions.
Among them, we manually confirm that $1,667$ ones (D) are false positives.

\begin{table}[t]
	\caption{Comparison between \tool{} and Etherscan in identifying \defi{} actions. }
	\label{tab:evaluate_action_identification}
	\centering
	
	\resizebox{0.7\textwidth}{!}
    {
    \begin{tabular}{lllllllll}
        \toprule
                            & \multicolumn{1}{l}{\begin{tabular}[c]{@{}c@{}}\textbf{\# Identified}\\\textbf{  actions}\end{tabular}}  
                            & \textbf{\# TP} & \textbf{\# FP} & \textbf{\# TN} & \textbf{\# FN} 
                            & \textbf{FNR} & \textbf{TPR} & \textbf{Accuracy} \\
        \midrule
        \textbf{Etherscan}  & $\textbf{29,201}$ & $\textbf{29,201}$ & $\textbf{0}$ & $\textbf{1,667}$ & $\textbf{10,634}$ & $\textbf{26.7\%}$ & $\textbf{100\%}$ & $\textbf{74.38\%}$ \\
        \textbf{\tool{}}    & $\textbf{37,732}$ & $\textbf{36,065}$ & $\textbf{1,667}$ & $\textbf{0}$ & $\textbf{3,770}$ & $\textbf{9.46\%}$ & $\textbf{95.58\%}$ & $\textbf{86.9\%}$ \\
        \bottomrule
 	\end{tabular}}
\end{table}





\begin{equation}
    \label{formula:accuracy_precision}
    Accuracy = \frac{TP + TN}{TP + FP + TN + FN} * 100\%
\end{equation}

Table~\ref{tab:evaluate_action_identification} shows the detailed result. 
Since we do not have the ground truth, all actions identified by
Etherscan are treated as true positives (\textit{this assumption actually favors Etherscan}).
Then we use formula~(\ref{formula:accuracy_precision}) to compute the accuracy. 
Specifically, the accuracy of \tool{} ($86.9\%$) is higher than Etherscan ($74.38\%$).
What's more important, the false negative rate of \tool{} ($9.46\%$) is much lower than
Etherscan ($26.7\%$). This means \tool{} can identify more actions that will be missed
by  Etherscan. Since the attack detection is based on identified DeFi actions,
the miss of DeFi actions will lead to the miss of real-world attacks.
Until the writing of this paper in April $2021$,
Etherscan can identify none of the DeFi actions {for victim \defi{} apps} that are involved in the reported attacks by
\tool{} (Section~\ref{subsec:evaluate_effectiveness} {and Section~\ref{sec:forensic_analysis}}).

We summarize the advantages and disadvantages of our method to identify DeFi actions in the following.

\smallskip
\noindent\textbf{Advantages}\tab
Our method to identify \defi{} actions is
\textit{platform-independent}, which saves the manual efforts to understand
every new \defi{} protocol. Note that, DeFi apps tend to have
their own protocols.
Moreover, our system can detect more DeFi actions
than Etherscan, which leads to the detection of real attacks (Section~\ref{subsec:evaluate_effectiveness}).

\smallskip
\noindent\textbf{Disadvantages}\tab
	\tool{} misidentifies actions in a few cases, such as the exchange of more than two
types of tokens and adding multiple types of  tokens to provide liquidity.
This causes $1,667$ incorrect DeFi actions (with a false positive rate of $4.42\%$).
Our system can tolerate such
false positives since the detected attacks will be manually confirmed.

\begin{framed}
	\noindent \textbf{Summary:}
	\textit{
		\tool{} can identify more DeFi actions than Etherscan, with a lower
		false negative rate. The identified DeFi actions are critical to detect
		price manipulation attacks reported in Section~\ref{subsec:evaluate_effectiveness} and analyzed in Section~\ref{sec:forensic_analysis}}.
\end{framed}

    



\begin{table*}[t]
	\centering
    \caption{The overview of transactions involved in price manipulation attacks.}
    \resizebox{1.0\textwidth}{!}{
	\begin{tabular}{lccccccc}
    \toprule
        & \textbf{\# Total}      & \textbf{ \# Labeled }  & \multicolumn{2}{c}{ \textbf{True positives} }               & \multicolumn{2}{c}{ \textbf{False positives} }                    & \textbf{ \# Unconfirmed}  \\
        & \textbf{transactions}  & \textbf{transactions } & \textbf{\# Successful Attacks} & \textbf{\# Failed Attacks} & \textbf{\# Arbitrages} & \textbf{\# Recognition errors$^{\rm a}$} & \textbf{transactions } \\
    \hline
    \textbf{Indirect \pricemani{} attack} & \multirow{2}{*}{ $\textbf{350,823,625}$ } & $\textbf{101}$ & $\textbf{41}$ & $\textbf{3}$ & $\textbf{51}$ & $\textbf{0}$  & $\textbf{6}$ \\
    \textbf{Direct \pricemani{} attack}   &                                 & $\textbf{423}$ & $\textbf{391}$ & $\textbf{0}$ & $\textbf{0}$  & $\textbf{31}$ & $\textbf{1}$ \\
    \bottomrule
    \end{tabular}}
    \begin{tablenotes}
        \item $\rm a$:  The recognition error represents false positives caused by misidentified \defi{} actions.
    \end{tablenotes}

\label{tab:effectiveness}
\end{table*}


\subsection{Price Manipulation Attack Detection}
\label{subsec:evaluate_effectiveness}
We performed a large-scale detection on the $350,823,625$ transactions from Nov. 25, 2019, to Dec. 17, 2020.
As shown in Table~\ref{tab:effectiveness}, \tool{} detected $524$ attacks in total. We manually confirmed that
$432$ of them are true positives. In total, nine security incidents are detected. Four of them are known~\cite{Harvest-hack, Cheese-bank-incident, Value-incident, WarpFinance-incident}, while five of them
are \textit{zero-day} incidents. We reported our findings and two CVEs were assigned. 

The false positives consist of $51$ transactions for arbitrage and $31$ transactions with
misidentified \defi{} actions. 
First, our arbitrage detection strategy in Section~\ref{subsec:detection} misses 
$51$ arbitrage transactions. This can be improved with a more accurate pattern to identify
arbitrage. Second, \tool{} misidentifies a borrow action of bZx~\cite{bZx} as a {trade}.
As a result, it further misidentifies a borrow-redeem pair as a pair of reverse {trades},
which causes the false positives. That's because \tool{} does not yet support the identification
of loan-related actions.

In summary, \tool{} not only detects  four known security 
incidents~\cite{Harvest-hack, Cheese-bank-incident, Value-incident, WarpFinance-incident} but also reveals 
five \textit{zero-day} incidents.
In addition, \tool{} is the \textit{first} tool to detect price manipulation attacks in 
Ethereum. 
Besides, \tool{} finishes the large-scale experiment in $10$ hours and 
consumes negligible memory during the process.

\begin{framed}
	\noindent \textbf{Summary:}
	\textit{The detected nine security incidents with $432$ attacks shows the effectiveness of \tool{}. Among them,
	five security incidents are zero-day ones (with $399$ attacks in the wild.) 
	To the best of our knowledge, \tool{} is the \textit{first}
	tool that has the capability to detect price manipulation attacks in 
	Ethereum.}
\end{framed}

	\section{Attack Analysis}
\label{sec:forensic_analysis}

\begin{table*}[t]
	\centering
	\caption{The overview of security incidents detected by \tool{}.}
    \small
	\resizebox{1.06\textwidth}{!}{
		\hspace{-0.6in}
	\begin{tabular}{llllllll}
    \toprule
        \multicolumn{8}{l}{\textbf{Direct \pricemani{} attacks}} \\
        \multicolumn{1}{c}{\begin{tabular}[c]{@{}c@{}}\textbf{Date(s)}\\\textbf{(all in 2020)}\end{tabular}} & \textbf{Incident} & \textbf{\# Attacks} & \textbf{Profit} & \textbf{Victim app's type} & \textbf{Root cause} & \textbf{CVE-ID} & \textbf{Zero-day?} \\
    \midrule
        \code{Sep 30, 20 - Oct 15, 20}     & LRC Incident      & $91$  & $80.97$ ETH $^{\rm a}$     & DEX        & Insufficient access control     & \textit{CVE-2021-3xxx} & $\checkmark$\\
        \code{Oct 08, 20 - Dec 11, 20}     & DRC Incident      & $78$ & $25.05$ ETH $^{\rm b}$      & Portfolio management  & Insufficient access control     & - & $\checkmark$ \\
        \code{Nov 15, 20 - Dec 15, 20}     & SEAL Incident     & $24$  & $4.247K$ SEAL $^{\rm c}$   & Portfolio management  & Insufficient access control    & \textit{CVE-2021-3xxx} & $\checkmark$\\
        \code{Nov 26, 19 - Oct 24, 20}     & MET Incident & $198$ & $1.16$ ETH $^{\rm d}$          & Auction   & Insufficient access control     & -                      & $\checkmark$ \\
    \midrule
        \multicolumn{8}{l}{\textbf{Indirect \pricemani{} attacks}} \\
    \midrule
        \code{Oct 26, 20} & Harvest Hack~\cite{Harvest-hack}                   & $30$& $24M$ stablecoins of USD   & Portfolio management & \multicolumn{2}{l}{ Insecure price dependency } & $\times$ \\
        \code{Oct 29, 20} & Plouto Hack                                        & $8$ & $650K$ stablecoins of USD  & Portfolio management & \multicolumn{2}{l}{ Insecure price dependency } & $\checkmark$ \\
        \code{Nov 06, 20}  & Cheese Bank Incident~\cite{Cheese-bank-incident}   & $1$ & $3.3M$ stablecoins of USD  & Lending       & \multicolumn{2}{l}{ Insecure price dependency } & $\times$ \\
        \code{Nov 14, 20} & Value DeFi Incident~\cite{Value-incident}          & $1$ & $7.4M$ stablecoins of USD  & Portfolio management & \multicolumn{2}{l}{ Insecure price dependency } & $\times$ \\
        \code{Dec 17, 20} & WarpFinance Incident~\cite{WarpFinance-incident}   & $1$ & $3.81M$ stablecoins of USD & Lending       & \multicolumn{2}{l}{ Insecure price dependency } & $\times$ \\
    \bottomrule
	\end{tabular}}
\label{tab:result_overview}
    \begin{tablenotes}
        \item $\rm a$: $80.97$ ETH was worth around $29K$ USD on Sep 30; $\rm b$: $25.05$ ETH was worth around $8.5K$ USD on Oct 8;
        \item $\rm c$: $4.247K$ SEAL was worth around $306K$ USD on Nov 15; $\rm d$: $1.16$ ETH was worth around $603$ USD on Nov 26.
    \end{tablenotes}

\end{table*}

In this section, we present the analysis result on the
confirmed security incidents
and insights to secure the \defi{} ecosystem.

\subsection{Vulnerability Root Cause Analysis}
\label{subsec:vulnerabilities}
After analyzing each security incident, we find that
the root causes of the vulnerabilities are falling
into two categories, i.e., insufficient access control
of {the smart contract's APIs}
and insecure price dependency between DeFi apps. 

\subsubsection{Insufficient access control}
Four DeFi apps, including Loopring~\cite{Loopring},
Dracula~\cite{Dracula}, Seal Finance~\cite{SealFinance}, and Metronome~\cite{Metronome},
are susceptible to {direct} \pricemani{} attacks.
The root cause of the vulnerability is the same: 
they do not properly enforce the access control to the function that may change the AMM's reserves.

The four vulnerable functions are \code{sellTokenForLRC}, \code{drain},
\code{breed} and \code{closeAuction} in these apps.
The first three could be exploited by an attacker to
sell cryptocurrencies in  a liquidity pool of an AMM.
An attacker can manipulate the pool in advance to decrease related 
cryptocurrencies' prices and then invoke these functions to sell them \textit{cheaply},
which causes the vulnerable apps to suffer losses.
After that, the attacker can make profits by buying these
cryptocurrencies at a relatively low price. 
The last one (\code{closeAuction}) directly deposits Ether into the  Metronome's AMM pool.
Thus, an attacker first swaps Ether for the MET token and then invokes
\code{closeAuction} to increase the AMM's Ether reserves, which lifts MET's price.
Finally, the attacker swaps the  same amount of  MET for more
Ethers than that are used in the first swap.
As a result, the attacker can make profits   with each invocation of \code{clouseAuction}.

Note that, Dracula~\cite{Dracula}  implemented 
an access control for the function \code{drain} in the latest version~\cite{Dracula-patch}.

\begin{framed}
    \noindent \textbf{Insight I:}
    \textit{
        {Insufficient} access control for functions that can change the AMM's reserves makes the vulnerable DeFi
        apps susceptible to direct \pricemani{} attacks.  }
\end{framed}

\begin{table}[t]
	\centering
    \caption{Symbols used in formula~(\ref{formula:harvest_pricing}),
    	(\ref{formula:cheese_pricing}), and (\ref{formula:warp_pricing}).}
    \resizebox{0.48\textwidth}{!}{
    	\hspace{-0.2in}
	\begin{tabular}{ll}
	\toprule
    \textbf{Symbol} & \textbf{Description} \\
    \midrule
        $AM_{asset}$ & The amount of the $asset$ minted to a user. \\
        $AB_{asset}$ & The amount of the $asset$ burning by a user. \\
        $AD_{asset}$ & The amount of the $asset$ depositing by a  user. \\
        $AR_{asset}$ & The amount of the $asset$ redeemed to a  user. \\
        $R_{asset}$ & The current reserves of the $asset$ in the pool. \\
        $TS_{asset}$ & The total supply of the $asset$. \\
        $UP_{asset}$ & The unit price of the $asset$. \\
    \bottomrule
	\end{tabular}}
\label{tab:symbols}
\end{table}

\subsubsection{Insecure price dependency}
\label{subsubsec:insecure_price_depedency}
Two types of vulnerable apps are susceptible to the {indirect} \pricemani{} attack, including
portfolio management apps and lending apps.
The portfolio management apps \textit{leverage AMM's real-time quotation to price clients' deposits}, and lending
apps \textit{rely on AMM's real-time reserves to price clients' collaterals.}
Thus, an attacker can manipulate the real-time quotation and reserves in AMMs
to attack vulnerable apps.

Table~\ref{tab:symbols} shows the symbols that will be used in the formula~(\ref{formula:harvest_pricing}),
formula~(\ref{formula:cheese_pricing}), and formula~(\ref{formula:warp_pricing}).

\smallskip \noindent \textbf{Depending on AMM's real-time quotation}\tab
Harvest~\cite{Value-DeFi}, Plouto~\cite{Plouto}, and Value DeFi~\cite{Value-DeFi} are portfolio management apps. 
They provide the same financial services and share a similar price mechanism. We use Harvest as a
representative example.

The attack flow of the Harvest Hack~\cite{Harvest-hack} has been described
in Section~\ref{subsec:show_example}. We focus on its insecure price dependency. 
After receiving a user's stablecoins, such as USDC, 
Harvest automatically deposits stablecoins to a DeFi app  
with the highest Annual Percentage Yield (APY).
In short, Harvest acts as an account manager that helps users (clients) to invest their stablecoins.

One of Harvest's investment strategies is to deposit clients' USDC into
Curve Y pool~\cite{curveYpool}  to provide liquidity. 
During the process, clients get fUSDC (Harvest's LP token) from Harvest as certificates, and Harvest gets 
yCrv (Y pool's LP token) from Y pool as certificates. 
Formula~\ref{formula:harvest_pricing} describes the number of fUSDC tokens that Harvest should mint and send
to clients after receiving clients' deposits (USDC), and the number of USDC that Harvest should refund to
clients after receiving and burning clients' fUSDC.

\begin{equation}
    \label{formula:harvest_pricing}
    \begin{aligned}
    & AM_{fUSDC} = \frac{AD_{USDC}}{R_{USDC} + \bm{ToUSDC}(R_{yCrv})} * {TS_{fUSDC}} \\
    & AR_{USDC} = \frac{AB_{fUSDC}}{TS_{fUSDC}} * (R_{USDC} + \bm{ToUSDC}(R_{yCrv})) \\
    \end{aligned}
\end{equation}

According to the formula~(\ref{formula:harvest_pricing}), the amount of minted fUSDC equals the product of the   
deposited USDC's share and the total supply of fUSDC. The amount of refunded USDC equals the product of burned
fUSDC's share and the Total Value Locked (TVL). Since Harvest invests part of USDC into Curve, the TVL consists of the 
reserves of USDC and yCrv. To unify the unit, Harvest leverages the \textit{real-time} price provided by the Curve Y
pool ($ToUSDC$). However, the \textit{real-time} price of the AMM (Curve) can be manipulated.
Therefore, attackers can make profits  through increasing the amount of refunded USDC by manipulating the Curve Y pool,
as shown in the Harvest Hack Section (Section~\ref{subsec:show_example}).

To the best of our knowledge, using the average price over a period of time can mitigate the
{indirect} \pricemani{} attack. In fact, Curve has the interface (\code{virtual\_price})
that provides the average token price. However, Harvest does not use this interface in its smart contract.

\smallskip \noindent \textbf{Depending on AMM's real-time reserves}\tab
Cheese Bank~\cite{Cheese-Bank} and WarpFinance~\cite{WarpFinance} are  lending apps. They support
Uniswap V2's LP token as the collateral. Since Uniswap does not provide
its LP tokens' price, two lending
apps implement their own pricing mechanism for Uniswap's LP tokens.

\begin{equation}
    \label{formula:cheese_pricing}
    UP_{LP token} = \frac{R_{ETH} * 2 * UP_{ETH}}{TS_{LP token}}
\end{equation}

Each Uniswap V2 pool maintains a pair of cryptocurrencies. 
Cheese Bank only supports the LP tokens of a limited number of Ether-related pools, e.g,
Ether-USDT, Ether-USDC, Ether-CHEESE, and 
Ether-DAI. As shown in formula~(\ref{formula:cheese_pricing}), it takes twice the value of Ether reserves
in the pool as the pool's TVL, which decides the LP token's unit price.

\begin{equation}
    \label{formula:warp_pricing}
    UP_{LP token} = \frac{UP_{token0} * R_{token0} + UP_{token1} * R_{token1}}{TS_{LP token}}
\end{equation}

Formula~(\ref{formula:warp_pricing}) shows how WarpFinance prices Uniswap V2's LP token. Differ from Cheese Bank, 
WarpFinance takes the value of all reserves in an Uniswap V2's pool as the pool's TVL. Then, it takes a pool's
TVL divided by the total supply of the pool's LP token as the unit price of the LP token.

The two price mechanisms are vulnerable because they both depend on the \textit{real-time}
reserves of Uniswap V2's pools ($R_{ETH}$, $R_{token0}$, and $R_{token1}$).
Therefore, attackers can lift the collateral's price by manipulating the related Uniswap V2's pool in advance.
After collateralizing the ``expensive collaterals'', attackers can borrow out more valuable stablecoins. They will, of course, not redeem their collaterals.

\begin{framed}
    \noindent \textbf{Insight II:}
    \textit{
        The insecure price dependency is the root cause of vulnerable apps
        that are susceptible to indirect price manipulation attacks.
    }
\end{framed}

\begin{figure}[t]
	\centering
		\ifx\arxiv\undefined
\includegraphics[width=0.45\textwidth]{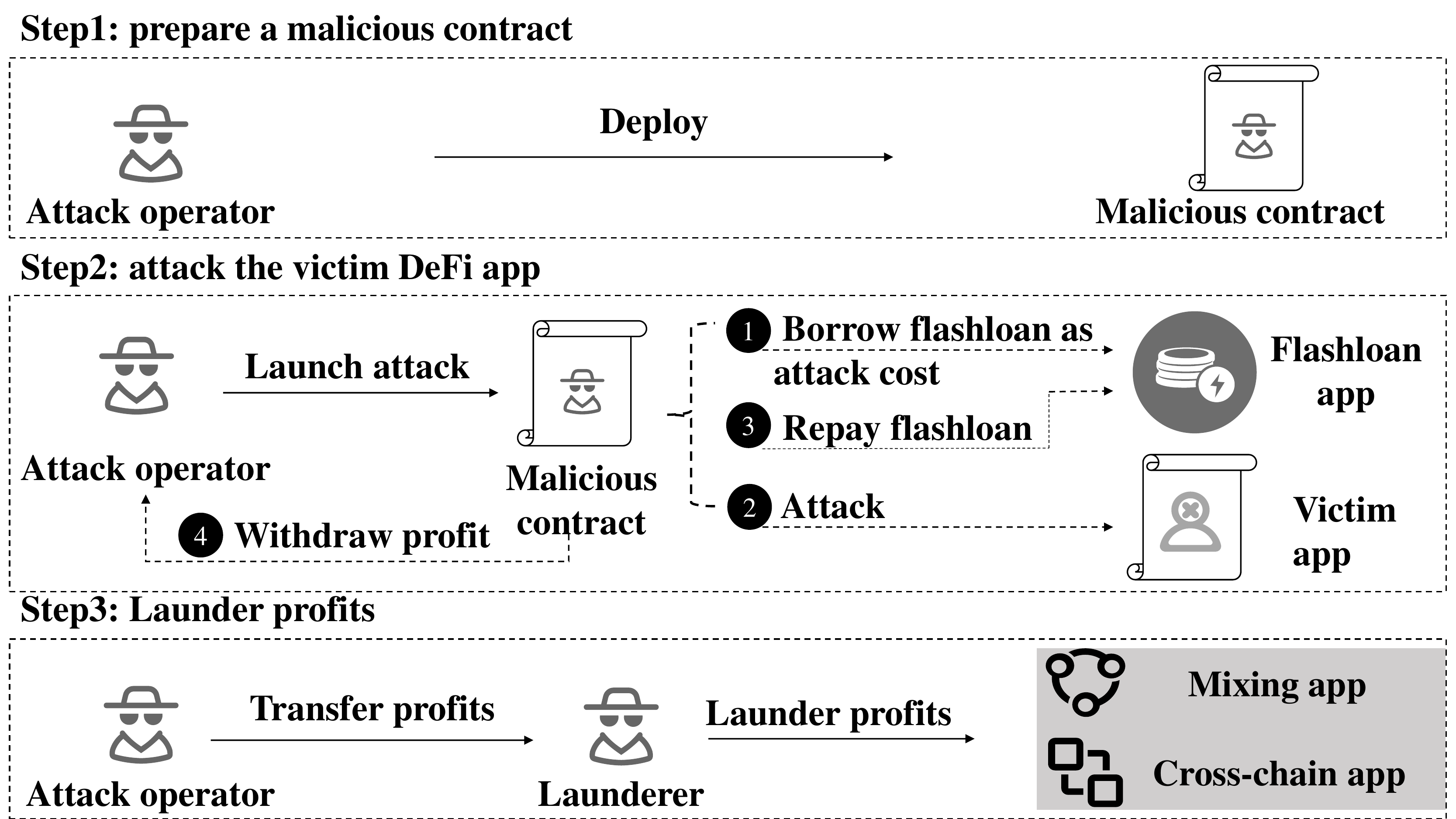}
	\else
\includegraphics[width=0.6\textwidth]{pdfs/attack_footprint.pdf}
	\fi

    \caption{Attack footprint. The attack operator and token launderer both are EOAs.}
	\label{fig:footprint}
\end{figure}

\subsection{Footprint analysis}

Attackers have used similar steps to launch the attack. We summarize the attack footprint
in Fig.~\ref{fig:footprint}.

\begin{itemize} [leftmargin=*]
	\item{\textit{Prepare a malicious contract}:} The attacker uses an EOA, i.e., \textit{the attack operator},
	to deploy a malicious contract.
	\item{\textit{Attack the vulnerable apps}:} \textit{The attack operator} invokes the malicious
	contract to launch the attack. Note that, the attacker usually leverages the flash loan
	to get a large number of tokens that are needed during the attack.
	\item{\textit{Launder profits}:} \textit{The attack operator} then transfers profits to
	another EOA, i.e., \textit{the token launderer}. The token launderer leverages the mixing application,
	such as tornoda~\cite{tornado}, or a cross-chain application, such as 
	Ren~\cite{Ren}, to launder the profits. 
\end{itemize}

Furthermore, we check whether a malicious contract has other callers or whether a malicious
contract's creator has deployed other smart contracts.
The result is interesting. There are almost no other transactions except for the ones that are used
for attacks. This \textit{clean} attack strategy  makes tracing attackers more challenging.

\begin{framed}
    \noindent \textbf{Insight III:}
    \textit{
    	 Attackers are becoming mature. They leverage the mixing service and
    	 cross-chain apps to launder the profits, and hide their traces by using a \textit{clean}
    	 attack strategy.
}
\end{framed}

\begin{figure}[t]
	\centering
	\ifx\arxiv\undefined
	\includegraphics[width=0.4\textwidth]{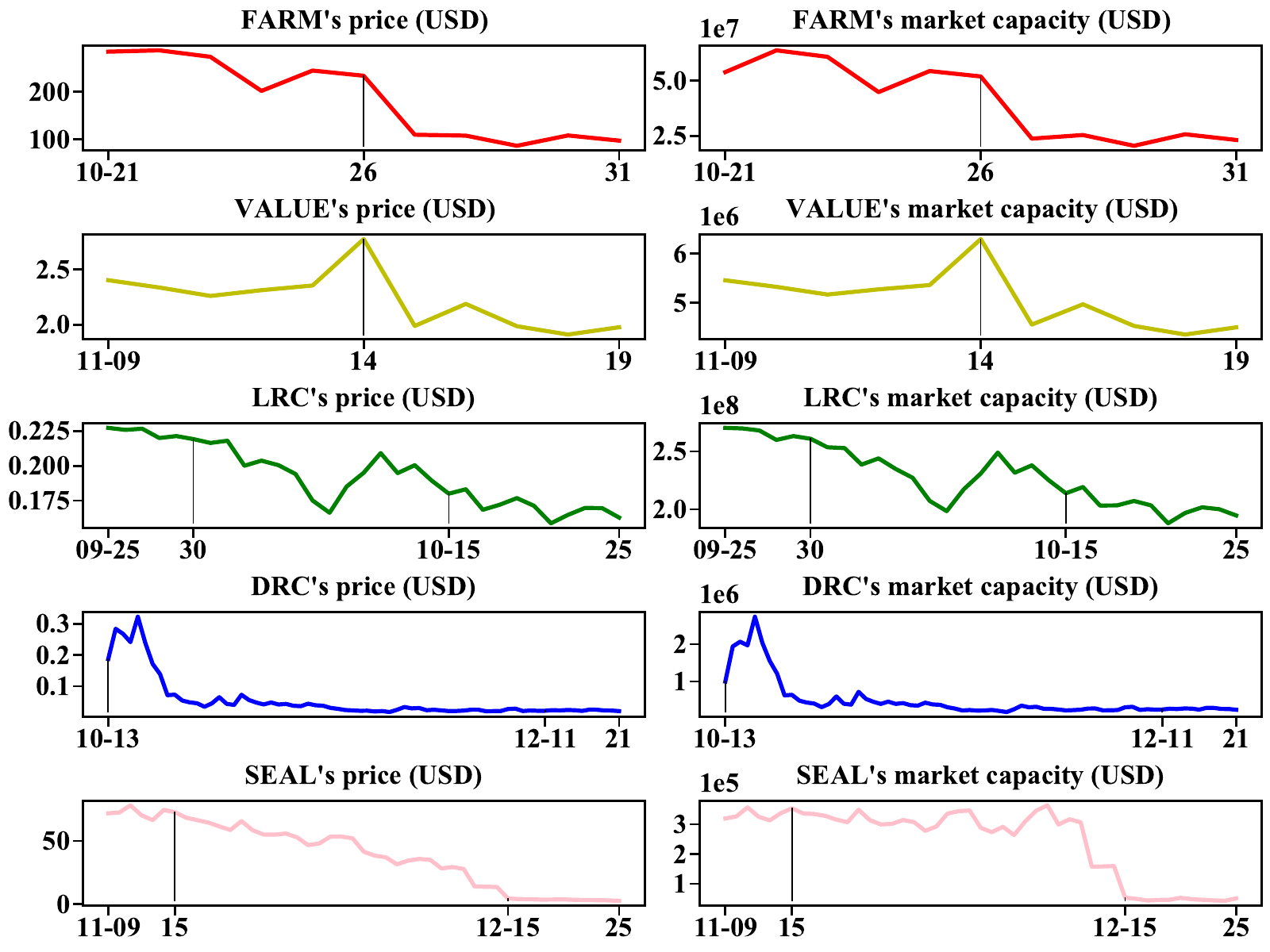}
	\else
	\includegraphics[width=0.6\textwidth]{pdfs/impacts_on_victims.pdf}
	\fi

    \caption{The market value trend of vulnerable DeFi applications around the day of the related security incidents. 
    The price is fetched from CoinGecko~\cite{coingecko}. It does not store the DRC's token price before Oct 13.
    Lines refer to the day of incidents shown in Table~\ref{tab:result_overview};
    FARM is the protocol token of Harvest.}
	\label{fig:victim_impact}
\end{figure}

\subsection{Impact Analysis}
\label{subsec:impact}

We analyze the impact on vulnerable DeFi apps' market value after the security incidents. 
As shown in Fig.~\ref{fig:victim_impact}, we use the unit price and market capacity (in USD) to 
describe the market value of five vulnerable apps' protocol tokens (tokens
issued by DeFi apps).
The price is retrieved from CoinGecko~\cite{coingecko}.

The market value of Harvest and Value dropped \textit{sharply} on the day of the related incidents.
We suspect that they are caused by the security incidents.
In addition, the market value of Loopring, Dracula, and Seal Finance is also falling after the day
of the related incidents.
However, we cannot simply conclude that they are affected by related security incidents due to  two main reasons.
First, the market values are declining before related incidents.  
Second, LRC incident, DRC incident, and SEAL incident are \textit{zero-day}. The lack of
visibility limits their follow-up impacts.

\begin{framed}
	\noindent \textbf{Insight IV:}
	\textit{
		Security incidents have direct impact on the market value of vulnerable DeFi apps, especially
		for the publicly known attacks.
	}
\end{framed}

	\section{Discussion}
\label{sec:discussion}

In this section, we discuss \tool{'s} limitations and future improvements.


\smallskip \noindent \textbf{Identify more \defi{} actions}\tab
Currently, \tool{} cannot identify a few \defi{} actions, such 
as loan-related actions. 
For example, among   $3,770$ missed
actions compared with Etherscan (Section~\ref{subsec:evaluation_action_indentification}),
$2,182$ of them are loan-related actions.
We will 
extend \tool{'s} capability to identify more \defi{} actions as one of the
future work.

\smallskip \noindent \textbf{Detect more types of \defi{} attacks}\tab
Currently, \tool{} only focuses on two types of \pricemani{}
attacks. In fact, with the recovered \defi{} semantics, 
\tool{} is able to detect more types of \defi{} attacks,
if the detection patterns can be developed.
To secure the Ethereum \defi{} ecosystem, 
we will design more attack patterns and extend 
\tool{'s} detection ability. 
This process might be coupled with the extension of \defi{} semantics identification.

	\section{Related Work}
\label{sec:related_work}

\subsection{ Ethereum \defi{} ecosystem }
The wave of \defi{} brings many traditional financial applications to Ethereum, which attempt to use the openness and 
transparency of the blockchain technology to evolve finance from opaque to transparency. Werner et al.~\cite{werner2021sok}
summarize  \defi{'s} current development, including \defi{} basic knowledge, popular  \defi{} protocols, and
\defi{} security issues. Particularly, Clark et al.~\cite{clark2019sok}, Pernice et al.~\cite{pernice2019monetary}, 
and Moin et al.~\cite{moin2019classification} perform the empirical study about current stablecoins' protocol design and 
provide the comprehensive taxonomy of stablecoins with their insights. Wang~\cite{wang2020automated} studies 
mathematical models of AMMs. Bartoletti et al.~\cite{bartoletti2020sok} study various implementations of existing lending 
apps. Furthermore, Perez et al.~\cite{perez2020liquidations} provide the first in-depth empirical study for liquidations
on protocols for loanable funds (PLFs), such as lending apps.  
Liu et al.~\cite{liu2020first} present the first measurement
for \defi{} price oracles. 
The knowledge of these studies can help better understand the DeFi ecosystem.

\subsection{\defi{} security }
\subsubsection{Off-chain security issues} 
Front-running and Pump-and-Dump are two typical threats for DeFi security. They mainly focus on are off-chain activities. 

\smallskip \noindent \textbf{Front-running}\tab
In traditional finance, front-running often refers to a broker prioritizing his trade ahead of his clients' market-moving 
order to benefit himself. 
Front-runners prioritize their transactions ahead of others 
by lifting transaction fees~\footnote{Miners tend to pack transactions with higher transaction fee.}. 
Eskandari et al.~\cite{eskandari2019sok} analyze the front-running issues across the top $25$ DeFi apps in Ethereum and 
present the evidence of abnormal miners' behaviors of purchasing cryptocurrencies for front-running. 
Daian et al.~\cite{daian2020flash} study bots' arbitrage strategies
and reveal the situation of front-running
between bots.
Zhou et al.~\cite{zhou2020highfrequency} formalize the sandwich attack combining  front- and back-running  actions and present an 
empirical evaluation on it.

\smallskip \noindent \textbf{Pump-and-Dump (P\&D)}\tab
P\&D often refers to a scheme that boosts the price of a stock by misleading information. 
The perpetrator of this scheme purchases a lot of these stocks in advance and sells them after the price has been driven up.
Recently, this type of scheme also appears on the Ethereum \defi{} ecosystem.  
Kamps et al.~\cite{kamps2018moon} construct a group of patterns by analyzing existing P\&D schemes and define a set of 
identifying criteria, which can detect suspected P\&D behaviors. 
Xu et al.~\cite{xu2019anatomy} aggregate $412$ P\&D schemes in telegram channels and build a machine learning model 
that can predict the likelihood of a cryptocurrency being pumped.

\subsubsection{On-chain security issues}
Various \defi{} protocols are arising but with many security issues. Particularly, Flash loan is a nascent service 
that can lend any unsecured cryptocurrencies to clients.
With this temporary
funding capacity, some security issues have surfaced. 
Gudgeon et al.~\cite{gudgeon2020decentralized} demonstrated a simulated governance attack with the flash loan on the 
MakerDao~\cite{maker}. 
Qin et al.~\cite{qin2020attacking} analyze two existing flash loan attacks and propose an optimization strategy that 
can increase the attacker's benefit to $2.37$ and $1.73$ times the previous. 
Wang et al.~\cite{wang2021blockeye} implement a preliminary work on automatically hunting for \defi{} attacks.

Our work is different from them and aims to detect two new types of DeFi attacks.

\subsection{Smart Contract Code Vulnerability}
Ethereum smart contracts suffer from code vulnerabilities,
such as re-entrancy and integer 
overflow. Many systems have been proposed to detect vulnerable
smart contracts~\cite{luu2016making, torres2018osiris, jiang2018contractfuzzer, wustholz2020harvey,  kalra2018zeus, tsankov2018securify, schneidewind2020ethor} or real-world attacks~\cite{grossman2017online, ferreira2020aegis, rodler2018sereum, chen2020soda, zhang2020txspector, Wu2020TimeTravelIT}.
For instance, Oyente~\cite{luu2016making}  applies the symbolic execution technique to detect code vulnerabilities. eThor~\cite{schneidewind2020ethor} leverages formal 
verification to reveal code vulnerabilities in smart contracts.
TXSPECTOR~\cite{zhang2020txspector} and EthScope~\cite{Wu2020TimeTravelIT} focus on uncovering historical 
attacks caused by code vulnerabilities in Ethereum. 
These systems cannot be directly applied to detect price manipulation attacks
since they lack the capability to recover DeFi semantics.

	\section{Conclusion}

In this work, we aim to detect two types of price manipulation attacks
on DeFi apps. To this end, we present a new approach to automatically
recover DeFi semantics from raw transactions.
Then we  detect  attacks based on the recovered DeFi semantics.
We implemented our approach in a tool named \tool{}.
The evaluation result shows that our system can accurately
recover \defi{} semantics, and effectively detect \defi{} attacks.
In total, it revealed $432$ real   real-world attacks that involve
four known security incidents and five
\textit{zero-day} security incidents.
Our work urges the need to
secure the \defi{} ecosystem.

	\bibliographystyle{plain}
	\bibliography{main}
\end{document}